\documentclass[twocolumn,aps,prc,superscriptaddress,amsmath,amssymb,floatfix]{revtex4}
\usepackage{graphicx}

\begin{document}

\title{Centrality and system size dependence of multiplicity fluctuations \\
in nuclear collisions at ${\bf 158\,{\it A}}$~GeV}

\affiliation{NIKHEF, Amsterdam, Netherlands.}
\affiliation{Department of Physics, University of Athens, Athens,
Greece.} \affiliation{Comenius University, Bratislava, Slovakia.}
\affiliation{KFKI Research Institute for Particle and Nuclear
Physics, Budapest, Hungary.} \affiliation{MIT, Cambridge, USA.}
\affiliation{Henryk Niewodniczanski Institute of Nuclear Physics,
Polish Academy of Sciences, Cracow, Poland.}
\affiliation{Gesellschaft f\"{u}r Schwerionenforschung (GSI),
Darmstadt, Germany.} \affiliation{Joint Institute for Nuclear
Research, Dubna, Russia.} \affiliation{Fachbereich Physik der
Universit\"{a}t, Frankfurt, Germany.} \affiliation{CERN, Geneva,
Switzerland.} \affiliation{Institute of Physics
\'Swi\c{e}tokrzyska Academy, Kielce, Poland.}
\affiliation{Fachbereich Physik der Universit\"{a}t, Marburg,
Germany.} \affiliation{Max-Planck-Institut f\"{u}r Physik, Munich,
Germany.} \affiliation{Charles University, Faculty of Mathematics
and Physics, Institute of Particle and Nuclear Physics, Prague,
Czech Republic.} \affiliation{Department of Physics, Pusan
National University, Pusan, Republic of Korea.}
\affiliation{Nuclear Physics Laboratory, University of Washington,
Seattle, WA, USA.} \affiliation{Atomic Physics Department, Sofia
University St. Kliment Ohridski, Sofia, Bulgaria.}
\affiliation{Institute for Nuclear Research and Nuclear Energy,
Sofia, Bulgaria.} \affiliation{Department of Chemistry, Stony
Brook Univ. (SUNYSB), Stony Brook, USA.} \affiliation{Institute
for Nuclear Studies, Warsaw, Poland.} \affiliation{Institute for
Experimental Physics, University of Warsaw, Warsaw, Poland.}
\affiliation{Faculty of Physics, Warsaw University of Technology,
Warsaw, Poland.} \affiliation{Rudjer Boskovic Institute, Zagreb,
Croatia.}

\author{C.~Alt}
\affiliation{Fachbereich Physik der Universit\"{a}t, Frankfurt,
Germany.}
\author{T.~Anticic}
\affiliation{Rudjer Boskovic Institute, Zagreb, Croatia.}
\author{B.~Baatar}
\affiliation{Joint Institute for Nuclear Research, Dubna, Russia.}
\author{D.~Barna}
\affiliation{KFKI Research Institute for Particle and Nuclear
Physics, Budapest, Hungary.}
\author{J.~Bartke}
\affiliation{Henryk Niewodniczanski Institute of Nuclear Physics,
Polish Academy of Sciences, Cracow, Poland.}
\author{L.~Betev}
\affiliation{CERN, Geneva, Switzerland.}
\author{H.~Bia{\l}\-kowska}
\affiliation{Institute for Nuclear Studies, Warsaw, Poland.}
\author{C.~Blume}
\affiliation{Fachbereich Physik der Universit\"{a}t, Frankfurt,
Germany.}
\author{B.~Boimska}
\affiliation{Institute for Nuclear Studies, Warsaw, Poland.}
\author{M.~Botje}
\affiliation{NIKHEF, Amsterdam, Netherlands.}
\author{J.~Bracinik}
\affiliation{Comenius University, Bratislava, Slovakia.}
\author{R.~Bramm}
\affiliation{Fachbereich Physik der Universit\"{a}t, Frankfurt,
Germany.}
\author{P.~Bun\v{c}i\'{c}}
\affiliation{CERN, Geneva, Switzerland.}
\author{V.~Cerny}
\affiliation{Comenius University, Bratislava, Slovakia.}
\author{P.~Christakoglou}
\affiliation{Department of Physics, University of Athens, Athens,
Greece.}
\author{P.~Chung}
\affiliation{Department of Chemistry, Stony Brook Univ. (SUNYSB),
Stony Brook, USA.}
\author{O.~Chvala}
\affiliation{Charles University, Faculty of Mathematics and
Physics, Institute of Particle and Nuclear Physics, Prague, Czech
Republic.}
\author{J.G.~Cramer}
\affiliation{Nuclear Physics Laboratory, University of Washington,
Seattle, WA, USA.}
\author{P.~Csat\'{o}}
\affiliation{KFKI Research Institute for Particle and Nuclear
Physics, Budapest, Hungary.}
\author{P.~Dinkelaker}
\affiliation{Fachbereich Physik der Universit\"{a}t, Frankfurt,
Germany.}
\author{V.~Eckardt}
\affiliation{Max-Planck-Institut f\"{u}r Physik, Munich, Germany.}
\author{D.~Flierl}
\affiliation{Fachbereich Physik der Universit\"{a}t, Frankfurt,
Germany.}
\author{Z.~Fodor}
\affiliation{KFKI Research Institute for Particle and Nuclear
Physics, Budapest, Hungary.}
\author{P.~Foka}
\affiliation{Gesellschaft f\"{u}r Schwerionenforschung (GSI),
Darmstadt, Germany.}
\author{V.~Friese}
\affiliation{Gesellschaft f\"{u}r Schwerionenforschung (GSI),
Darmstadt, Germany.}
\author{J.~G\'{a}l}
\affiliation{KFKI Research Institute for Particle and Nuclear
Physics, Budapest, Hungary.}
\author{M.~Ga\'zdzicki}
\affiliation{Fachbereich Physik der Universit\"{a}t, Frankfurt,
Germany.}\affiliation{Institute of Physics \'Swi\c{e}tokrzyska
Academy, Kielce, Poland.}
\author{V.~Genchev}
\affiliation{Institute for Nuclear Research and Nuclear Energy,
Sofia, Bulgaria.}
\author{G.~Georgopoulos}
\affiliation{Department of Physics, University of Athens, Athens,
Greece.}
\author{E.~G{\l}adysz}
\affiliation{Henryk Niewodniczanski Institute of Nuclear Physics,
Polish Academy of Sciences, Cracow, Poland.}
\author{K.~Grebieszkow}
\affiliation{Faculty of Physics, Warsaw University of Technology,
Warsaw, Poland.}
\author{S.~Hegyi}
\affiliation{KFKI Research Institute for Particle and Nuclear
Physics, Budapest, Hungary.}
\author{C.~H\"{o}hne}
\affiliation{Gesellschaft f\"{u}r Schwerionenforschung (GSI),
Darmstadt, Germany.}
\author{K.~Kadija}
\affiliation{Rudjer Boskovic Institute, Zagreb, Croatia.}
\author{A.~Karev}
\affiliation{Max-Planck-Institut f\"{u}r Physik, Munich, Germany.}
\author{D.~Kikola}
\affiliation{Faculty of Physics, Warsaw University of Technology,
Warsaw, Poland.}
\author{M.~Kliemant}
\affiliation{Fachbereich Physik der Universit\"{a}t, Frankfurt,
Germany.}
\author{S.~Kniege}
\affiliation{Fachbereich Physik der Universit\"{a}t, Frankfurt,
Germany.}
\author{V.I.~Kolesnikov}
\affiliation{Joint Institute for Nuclear Research, Dubna, Russia.}
\author{E.~Kornas}
\affiliation{Henryk Niewodniczanski Institute of Nuclear Physics,
Polish Academy of Sciences, Cracow, Poland.}
\author{R.~Korus}
\affiliation{Institute of Physics \'Swi\c{e}tokrzyska Academy,
Kielce, Poland.}
\author{M.~Kowalski}
\affiliation{Henryk Niewodniczanski Institute of Nuclear Physics,
Polish Academy of Sciences, Cracow, Poland.}
\author{I.~Kraus}
\affiliation{Gesellschaft f\"{u}r Schwerionenforschung (GSI),
Darmstadt, Germany.}
\author{M.~Kreps}
\affiliation{Comenius University, Bratislava, Slovakia.}
\author{A.~Laszlo}
\affiliation{KFKI Research Institute for Particle and Nuclear
Physics, Budapest, Hungary.}
\author{R.~Lacey}
\affiliation{Department of Chemistry, Stony Brook Univ. (SUNYSB),
Stony Brook, USA.}
\author{M.~van~Leeuwen}
\affiliation{NIKHEF, Amsterdam, Netherlands.}
\author{P.~L\'{e}vai}
\affiliation{KFKI Research Institute for Particle and Nuclear
Physics, Budapest, Hungary.}
\author{L.~Litov}
\affiliation{Atomic Physics Department, Sofia University St.
Kliment Ohridski, Sofia, Bulgaria.}
\author{B.~Lungwitz}
\affiliation{Fachbereich Physik der Universit\"{a}t, Frankfurt,
Germany.}
\author{M.~Makariev}
\affiliation{Atomic Physics Department, Sofia University St.
Kliment Ohridski, Sofia, Bulgaria.}
\author{A.I.~Malakhov}
\affiliation{Joint Institute for Nuclear Research, Dubna, Russia.}
\author{M.~Mateev}
\affiliation{Atomic Physics Department, Sofia University St.
Kliment Ohridski, Sofia, Bulgaria.}
\author{G.L.~Melkumov}
\affiliation{Joint Institute for Nuclear Research, Dubna, Russia.}
\author{A.~Mischke}
\affiliation{NIKHEF, Amsterdam, Netherlands.}
\author{M.~Mitrovski}
\affiliation{Fachbereich Physik der Universit\"{a}t, Frankfurt,
Germany.}
\author{J.~Moln\'{a}r}
\affiliation{KFKI Research Institute for Particle and Nuclear
Physics, Budapest, Hungary.}
\author{St.~Mr\'owczy\'nski}
\affiliation{Institute of Physics \'Swi\c{e}tokrzyska Academy,
Kielce, Poland.}
\author{V.~Nicolic}
\affiliation{Rudjer Boskovic Institute, Zagreb, Croatia.}
\author{G.~P\'{a}lla}
\affiliation{KFKI Research Institute for Particle and Nuclear
Physics, Budapest, Hungary.}
\author{A.D.~Panagiotou}
\affiliation{Department of Physics, University of Athens, Athens,
Greece.}
\author{D.~Panayotov}
\affiliation{Atomic Physics Department, Sofia University St.
Kliment Ohridski, Sofia, Bulgaria.}
\author{A.~Petridis}~\email[deceased]{}
\affiliation{Department of Physics, University of Athens, Athens,
Greece.}
\author{W.~Peryt}
\affiliation{Faculty of Physics, Warsaw University of Technology,
Warsaw, Poland.}
\author{M.~Pikna}
\affiliation{Comenius University, Bratislava, Slovakia.}
\author{J.~Pluta}
\affiliation{Faculty of Physics, Warsaw University of Technology,
Warsaw, Poland.}
\author{D.~Prindle}
\affiliation{Nuclear Physics Laboratory, University of Washington,
Seattle, WA, USA.}
\author{F.~P\"{u}hlhofer}
\affiliation{Fachbereich Physik der Universit\"{a}t, Marburg,
Germany.}
\author{R.~Renfordt}
\affiliation{Fachbereich Physik der Universit\"{a}t, Frankfurt,
Germany.}
\author{C.~Roland}
\affiliation{MIT, Cambridge, USA.}
\author{G.~Roland}
\affiliation{MIT, Cambridge, USA.}
\author{M. Rybczy\'nski}~\email[Corresponding author. E-mail address: ]{mryb@pu.kielce.pl}
\affiliation{Institute of Physics \'Swi\c{e}tokrzyska Academy,
Kielce, Poland.}
\author{A.~Rybicki}
\affiliation{Henryk Niewodniczanski Institute of Nuclear Physics,
Polish Academy of Sciences, Cracow, Poland.}
\author{A.~Sandoval}
\affiliation{Gesellschaft f\"{u}r Schwerionenforschung (GSI),
Darmstadt, Germany.}
\author{N.~Schmitz}
\affiliation{Max-Planck-Institut f\"{u}r Physik, Munich, Germany.}
\author{T.~Schuster}
\affiliation{Fachbereich Physik der Universit\"{a}t, Frankfurt,
Germany.}
\author{P.~Seyboth}
\affiliation{Max-Planck-Institut f\"{u}r Physik, Munich, Germany.}
\author{F.~Sikl\'{e}r}
\affiliation{KFKI Research Institute for Particle and Nuclear
Physics, Budapest, Hungary.}
\author{B.~Sitar}
\affiliation{Comenius University, Bratislava, Slovakia.}
\author{E.~Skrzypczak}
\affiliation{Institute for Experimental Physics, University of
Warsaw, Warsaw, Poland.}
\author{M.~Slodkowski}
\affiliation{Faculty of Physics, Warsaw University of Technology,
Warsaw, Poland.}
\author{G.~Stefanek}
\affiliation{Institute of Physics \'Swi\c{e}tokrzyska Academy,
Kielce, Poland.}
\author{R.~Stock}
\affiliation{Fachbereich Physik der Universit\"{a}t, Frankfurt,
Germany.}
\author{C.~Strabel}
\affiliation{Fachbereich Physik der Universit\"{a}t, Frankfurt,
Germany.}
\author{H.~Str\"{o}bele}
\affiliation{Fachbereich Physik der Universit\"{a}t, Frankfurt,
Germany.}
\author{T.~Susa}
\affiliation{Rudjer Boskovic Institute, Zagreb, Croatia.}
\author{I.~Szentp\'{e}tery}
\affiliation{KFKI Research Institute for Particle and Nuclear
Physics, Budapest, Hungary.}
\author{J.~Sziklai}
\affiliation{KFKI Research Institute for Particle and Nuclear
Physics, Budapest, Hungary.}
\author{M.~Szuba}
\affiliation{Faculty of Physics, Warsaw University of Technology,
Warsaw, Poland.}
\author{P.~Szymanski}
\affiliation{CERN, Geneva, Switzerland.}\affiliation{Institute for
Nuclear Studies, Warsaw, Poland.}
\author{V.~Trubnikov}
\affiliation{Institute for Nuclear Studies, Warsaw, Poland.}
\author{D.~Varga}
\affiliation{KFKI Research Institute for Particle and Nuclear
Physics, Budapest, Hungary.}\affiliation{CERN, Geneva,
Switzerland.}
\author{M.~Vassiliou}
\affiliation{Department of Physics, University of Athens, Athens,
Greece.}
\author{G.I.~Veres}
\affiliation{KFKI Research Institute for Particle and Nuclear
Physics, Budapest, Hungary.}\affiliation{MIT, Cambridge, USA.}
\author{G.~Vesztergombi}
\affiliation{KFKI Research Institute for Particle and Nuclear
Physics, Budapest, Hungary.}
\author{D.~Vrani\'{c}}
\affiliation{Gesellschaft f\"{u}r Schwerionenforschung (GSI),
Darmstadt, Germany.}
\author{A.~Wetzler}
\affiliation{Fachbereich Physik der Universit\"{a}t, Frankfurt,
Germany.}
\author{Z.~W{\l}odarczyk}
\affiliation{Institute of Physics \'Swi\c{e}tokrzyska Academy,
Kielce, Poland.}
\author{A.~Wojtaszek}
\affiliation{Institute of Physics \'Swi\c{e}tokrzyska Academy,
Kielce, Poland.}
\author{I.K.~Yoo}
\affiliation{Department of Physics, Pusan National University,
Pusan, Republic of Korea.}
\author{J.~Zim\'{a}nyi}~\email[deceased]{}
\affiliation{KFKI Research Institute for Particle and Nuclear
Physics, Budapest, Hungary.}

\collaboration{The NA49 Collaboration}

\date{\today}


\begin{abstract}

The centrality and system size dependence of multiplicity
fluctuations of charged particles produced in nuclear collisions
at $158\,A$~GeV was studied by the NA49 collaboration. Centrality
selected Pb+Pb collisions, semi-central C+C and Si+Si collisions
as well as inelastic p+p interactions were analyzed. The number of
projectile participants determined on an event-by-event basis was
used to characterize the collision centrality. The scaled variance
of the multiplicity distribution obtained in the forward rapidity
region ($1.1 <y_{c.m.} <2.6$) shows a significant increase towards
peripheral collisions. The results are similar for negatively and
positively charged particles and about $50\%$ larger for all
charged particles. String-hadronic models of nuclear reactions
without the fusion process do not reproduce the rise of
fluctuations from central towards peripheral collisions. The
measured centrality dependence can be reproduced in superposition
models with the assumption of contributions from target
participants to particle production in the forward hemisphere or
in string models with fusion.

\end{abstract}.

\maketitle


\section{Introduction}


Nucleus-nucleus collisions at relativistic energies have been
intensely studied over the last two decades. The main goal of
these efforts is to understand the properties of strongly
interacting matter under extreme conditions of high energy and
baryon densities for which the creation of a quark-gluon plasma
(QGP) is expected~\cite{Collins:1974ky,Shuryak:1980tp}. Various
collision characteristics and their collision energy dependence
suggest that a transient state of deconfined matter is created at
collision energies as low as $30\,A$~GeV~\cite{Gazdzicki:2004ef}.
Fluctuations of physical observables in heavy ion collisions have
been a topic of interest for some years as they may provide
important signals regarding the formation of a QGP. With the large
number of particles produced in heavy ion collisions at CERN SPS
and BNL RHIC energies it has now become feasible to study
fluctuations on an event-by-event basis~\cite{Heiselberg:2000fk}.
In a thermodynamical picture of the strongly interacting system
formed in the collision, the experimentally studied fluctuations
of particle multiplicities
\cite{Aggarwal:2001aa,Afanasev:2000fu,Roland:2004pu,Adams:2003st,Chai:2005fj},
mean transverse momenta
\cite{Anticic:2003fd,Adamova:2003pz,Adams:2003uw,Adler:2003xq},
and other global observables, are related to fundamental
properties of the system, such as specific heat
\cite{Stodolsky:1995ds,Stephanov:1999zu}, chemical potential, and
matter compressibility \cite{Mrowczynski:1997kz}. These, in turn,
may reveal information on the properties of the equation of state
near the QCD phase boundary
\cite{Stephanov:1998dy,Stephanov:1999zu,Gazdzicki:2003bb}.

The main objective of this work is to study how the multiplicity
fluctuations change with the varying number of nucleons
participating in the collision, with centrality and system size.
First results on the centrality dependence in Pb+Pb collisions at
$158\,A$~GeV published by the WA98
collaboration~\cite{Aggarwal:2001aa} indicated that the scaled
variance of the multiplicity distribution near midrapidity
increases towards peripheral collisions. Our analysis extends such
measurements to C and Si nuclei and carefully addresses the
contribution from fluctuations of the number of participants which
can dominate the fluctuations of extensive quantities such as the
produced particle multiplicity. Furthermore an extensive
comparison of the results to available models is performed. The
data sample consists of centrality tagged minimum bias Pb+Pb,
semi-central C+C and Si+Si as well as minimum bias p+p collisions
registered by the NA49 detector at the CERN SPS. The multiplicity
fluctuations were studied in the forward rapidity region ($1.1
<y_{c.m.} <2.6$). The collision centrality was characterized by
the number of projectile participants derived from the projectile
spectator energy measured in a forward calorimeter. Selection of
narrow intervals in this energy minimizes the variation of the
number of projectile participants. Although the number of
projectile and target participants are closely correlated, the
number of target participants cannot be tightly constrained with
this procedure. The implications of this problem will be discussed
in detail.

The paper is organized as follows. In Sec.~\ref{s:method} the
method of measuring multiplicity fluctuations is introduced and
discussed. The NA49 set-up is presented in
Sec.~\ref{s:experiment}. Experimental procedures, in particular
event and particle selection, detector acceptance and centrality
determination are discussed in Sec.~\ref{s:data}. The results on
centrality and system size dependence of multiplicity fluctuations
are presented in Sec.~\ref{s:results}. In
Sec.~\ref{s:extrapolation} our results are compared with those of
other experiments. Possible explanations of our results are
discussed in Sec.~\ref{s:models}. The paper closes with a summary
and conclusions.


\section{Multiplicity fluctuations}
\label{s:method}



\subsection{Observables}


Let $P(N)$ be the multiplicity distribution, then
\begin{equation}
\langle N\rangle= \sum N \, P(N)
\end{equation}
is the mean value of the distribution. The variance of the
multiplicity distribution is defined as
\begin{equation}
{\rm Var}(N) \equiv \sum (N - \langle N\rangle )^2
P(N)=\langle N^2\rangle - \langle N\rangle^2.
\end{equation}
Note, that for a Poisson distribution the variance equals the mean
value, ${\rm Var}\left(N\right)=\langle N\rangle$. Mean value and
variance of multiplicity distributions are the only observables
used in this analysis.


\subsection{Participants and spectators}


In a description of nuclear collisions the concept of participant
and spectator nucleons is very useful. The participants are
nucleons which are removed from the Fermi spheres of target or
projectile nuclei due to the collision. The remaining nucleons are
called spectators. In the case of central nucleus-nucleus
collisions, where the impact parameter $b$ is relatively small,
almost all nucleons participate in the collision. In particular,
the number of projectile participants, $N_P^{PROJ}$, approximately
equals the total number of projectile nucleons
$N_{P}^{PROJ}\approx A$. If the collision is peripheral (with
large impact parameter $b$) almost all nucleons are spectators,
$N_{P}^{PROJ}\ll A$. The number of projectile spectators
$N_{SPEC}^{PROJ}$ is given by $N_{SPEC}^{PROJ}=A - N_{P}^{PROJ}$.


\subsection{Multiplicity fluctuations in superposition models}


Superposition models of nuclear collisions, which are frequently
used, assume that secondary particles are emitted by independent
sources. A prominent example is the Wounded Nucleon Model
\cite{Bialas:1976ed}, in which the sources are wounded nucleons,
i.e. the nucleons that have interacted at least once. In this
model the number of participants would be equal to the number of
wounded nucleons.

In superposition models the total multiplicity is
given by
\begin{equation}
N=\sum_{i=1}^{N_S}m_i,
\end{equation}
where $N_S$ denotes the number of sources and $m_i$ is the
multiplicity of secondaries from the $i-$th source. When the
sources are identical and independent from each other, the mean
total multiplicity equals
\begin{equation}
\langle N \rangle=\langle N_S \rangle \langle m \rangle,
\label{nsr}
\end{equation}
where $\langle N_S \rangle$ is the mean number of sources and
$\langle m \rangle$ is the mean multiplicity from a single source.
The variance of the multiplicity distribution is
\begin{equation}
\label{var}
{\rm Var}(N)=\langle N_{S}\rangle {\rm Var}(m)
+\langle m\rangle^2 \, {\rm Var}(N_{S}),
\end{equation}
where ${\rm Var}(m)$ and ${\rm Var}(N_{S})$ denote the variance
of the distribution of single-source multiplicity and the
variance of the distribution of source number, respectively.

The scaled variance of the multiplicity distribution
${\rm Var}(N)/\langle N\rangle$ is a useful measure of
the multiplicity fluctuations. From Eqs.~(\ref{nsr}) and
(\ref{var}) one gets
\begin{equation}
\label{sc-var}
\frac{{\rm Var}(N)}{\langle N\rangle}=
\frac{{\rm Var}(m)}{\langle m \rangle}
+ \langle m\rangle \frac{{\rm Var}(N_S)}{\langle N_S \rangle}.
\end{equation}
Thus, in superposition models the measured scaled variance of the
multiplicity distribution is the sum of two contributions. The
first one describes the multiplicity fluctuations from a single
source while the second one accounts for the fluctuations of the
number of sources. To infer the fluctuations of main interest
given by the first term in Eq.~(\ref{sc-var}) we try to minimize
the contribution of the second term. In this experiment events
with a fixed number of projectile participants can be selected in
which, however, the number of target participants can still
fluctuate. The effect of the fluctuation of target participant
number at fixed number of projectile participants was
theoretically studied in~\cite{Gazdzicki:2005rr}. Multiplicity
fluctuations are measured in a forward rapidity window in which
one mainly expects particles produced from projectile
participants. Nevertheless the effect of remaining fluctuations of
the number of target participants can remain important, if
participants produce particles over a wide rapidity range. We
return to this point in Secs.~\ref{s:scaled} and \ref{s:models}.

\begin{figure*}
\begin{center}
\centerline{\includegraphics*[width=14cm]{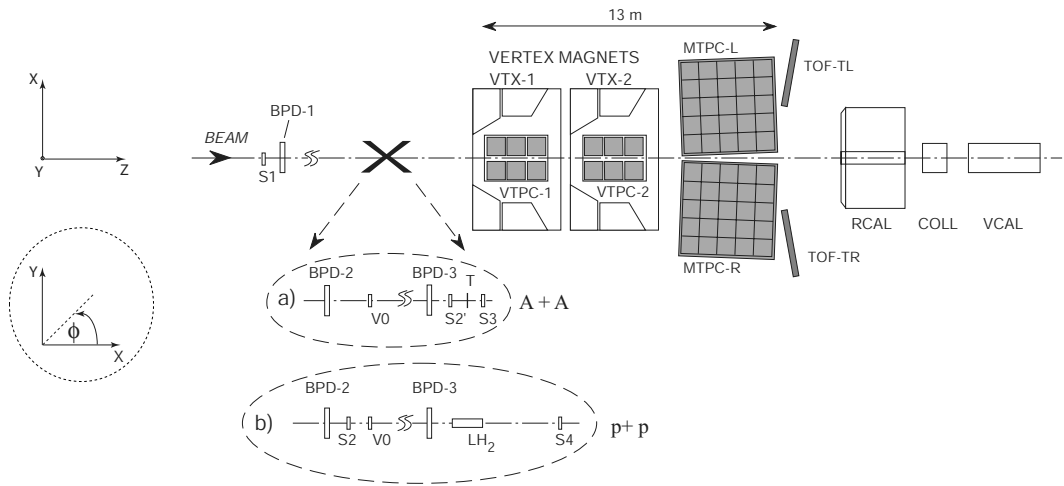}}
\caption{\label{setup}The experimental set-up of the NA49
experiment with different beam definitions and target
arrangements.}
\end{center}
\end{figure*}


\section{NA49 experimental set-up}
\label{s:experiment}


The NA49 experiment is a large acceptance hadron spectrometer at
the CERN-SPS (see Fig.~\ref{setup} and~\cite{na49_nim}) used to
study the final hadronic states produced in collisions of beam
particles (Pb directly from the SPS, C and Si via fragmentation of
the primary Pb beam, p as secondary beam from the 450 GeV proton
beam in the SPS) with a variety of fixed targets. The main
tracking devices are four large volume Time Projection Chambers
(TPCs) which are capable of detecting 60\% of some 1500 charged
particles created in a central Pb+Pb collision at $158\,A$~GeV.
Two of them, the Vertex TPCs (VTPC-1 and VTPC-2), are located
inside the magnetic field of two super-conducting dipole magnets
($1.5$ and $1.1$ T, respectively) and two others (MTPC-L and
MTPC-R) are positioned downstream of the magnets symmetrically to
the beam line. The results presented here are analysed with a
global tracking scheme~\cite{na49_global}, which combines track
segments that belong to the same physical particle but were
detected in different TPCs. The NA49 TPCs allow precise
measurements of particle momenta $p$ with a resolution of
$\sigma(p)/p^2 \cong (0.3-7)\cdot10^{-4}$ (GeV/c)$^{-1}$. The
set--up is supplemented by two Time of Flight (TOF) detector
arrays and a set of calorimeters.

The targets - C (561 mg/cm$^{2}$), Si (1170 mg/cm$^{2}$) discs and
a Pb (224 mg/cm$^{2}$) foil for ion collisions and a liquid
hydrogen cylinder (length 20.29 cm) for elementary interactions -
are positioned about 80 cm upstream from the VTPC-1.

Pb beam particles are identified by means of their charge as seen
by a Helium Gas-Cherenkov counter (S2') in front of the target.
The protons were identified by Cherenkov counters farther upstream
of the target. The study of C+C and Si+Si reactions is possible
through the generation of a secondary fragmentation beam which is
produced by a primary target (1 cm carbon) in the extracted
Pb-beam. Setting the beam line momentum accordingly, a large
fraction of all $Z/A = 1/2$ fragments are transported to the NA49
experiment. On-line selection based on a pulse height measurement
in a scintillator beam counter (S2) is used to select particles
with $Z=6,7$ (C, N) and $Z=13,14,15$ (Al, Si, P). Off-line
clean-up is achieved by using in addition the energy loss measured
by beam position detectors (BPD-1/2/3 in Fig.~\ref{setup}). These
detectors consist of pairs of proportional chambers and are placed
along the beam line for a precise measurement of the transverse
positions of the incoming beam particles.

For p beams, interactions in the target are selected by
anti-coincidence of the incoming beam particle with a small
scintillation counter (S4) placed on the beam trajectory between
the two vertex magnets. For p+p interactions at $158\,A$~GeV this
counter selects a (trigger) cross section of 28.23 mb out of 31.78
mb of the total inelastic cross section~\cite{pp_inelastic}. For
Pb-ion beams an interaction trigger is provided by
anti-coincidence with a Helium Gas-Cherenkov counter (S3) directly
behind the target. The S3 counter is used to select minimum bias
collisions by requiring a reduction of the Cherenkov signal by a
factor of about 6. Since the Cherenkov signal is proportional to
$Z^2$, this requirement ensures that the Pb projectile has
interacted, with a minimal constraint on the type of interaction.
This set-up limits the triggers on non-target interactions to rare
beam-gas collisions, the fraction of which proved to be small
after cuts, even in the case of peripheral Pb+Pb collisions.

The centrality of A+A collisions is selected by using information
from a Veto Calorimeter (VCAL), which measures the energy of the
projectile spectator nucleons. The geometrical acceptance of the
Veto Calorimeter is adjusted in order to cover the projectile
spectator region by a proper setting of the collimator (COLL).

Details of the NA49 detector set-up and performance of tracking
software are described in~\cite{na49_nim}.


\section{Data selection and analysis}
\label{s:data}


\subsection{Data sets}

Multiplicity fluctuations are studied for negatively, positively
and all charged particles selecting events within narrow intervals
of energy measured by the VCAL (predominantly energy of projectile
spectators). The experimental material used for the analysis
consists of samples of p+p, C+C, Si+Si and Pb+Pb collisions at
$158\,A$~GeV. The number of events in each sample is given in
Tab.~\ref{no_ev}. For Pb+Pb interactions a minimum bias trigger
was used allowing a study of centrality dependence.

\begin{table}
\caption{\label{no_ev}The number of events and the fraction of the
total inelastic cross section selected by the on-line trigger for
data sets used in this analysis.}
\begin{ruledtabular}
\begin{tabular}{lcr}
Data Set & No of events & $\sigma/\sigma^{inel}$ \\
\hline
p+p & 319 000 & 0.9\\
C+C & 51 000 & 0.153\\
Si+Si & 59 000 & 0.122\\
Pb+Pb & 165 000 & 0.6\\
\end{tabular}
\end{ruledtabular}
\end{table}


\subsection{NA49 acceptance}

\begin{figure}
\begin{center}
\centerline{\includegraphics*[width=8.5cm]{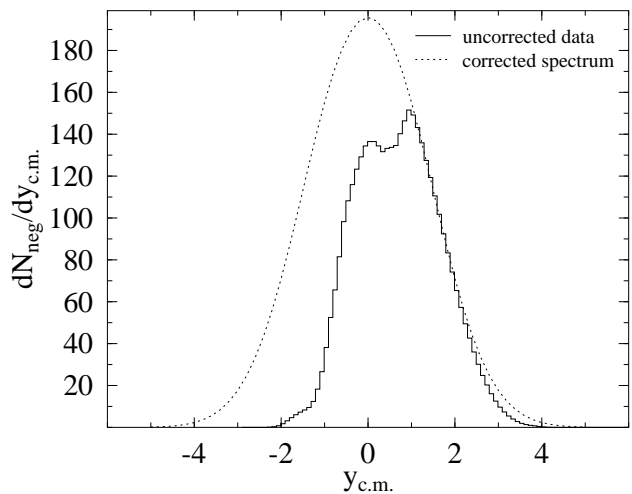}}
\caption{\label{y160a}Rapidity distribution of negatively charged
particles produced in central Pb+Pb collisions at $158\,A$~GeV
within the NA49 acceptance (full line) in comparison with an
acceptance corrected distribution from~\cite{Afanasiev:2002mx}
(dotted line).}
\end{center}
\end{figure}

The NA49 apparatus detects mainly the particles produced in the
forward rapidity hemisphere. Fig.~\ref{y160a} shows the rapidity
distribution of negatively charged particles in the NA49
acceptance calculated assuming that all particles are pions. This
distribution is compared to an acceptance corrected rapidity
distribution of negatively charged particles
from~\cite{Afanasiev:2002mx}. As seen, the rapidity region where
almost all particles are measured is $1.1 <y_{c.m.} <2.6$.

The NA49 detector was designed for a large acceptance in the
forward hemisphere. However, also in this region the geometrical
acceptance is not complete. The acceptance limits in the transverse
momentum and azimuthal angle were parameterized by a simple function
\begin{equation}
p_T(\phi) = \frac{1}{A+\frac{\phi^2}{C}} + B,
\label{eq_acc}
\end{equation}
where the values of $A$, $B$ and $C$ depend on the rapidity
interval, see~\cite{Anticic:2003fd}. Only particles within the
curves are used in this analysis. This well defined acceptance is
essential for later comparison of the results with models and
other experiments. Only forward rapidity tracks ($1.1 <y_{c.m.}
<2.6$, rapidity calculated assuming pion mass for all particles)
with $0.005< p_T < 1.5$~GeV/c have been used in this analysis.


\subsection{Event and particle selection}


The aim of the event selection criteria is to reduce contamination
from non-target collisions. The primary vertex was reconstructed
by fitting the intersection point of the measured particle
trajectories. Only events with a proper quality and position of
the reconstructed vertex were accepted for further analysis. The
vertex coordinate $Z$ along the beam had to satisfy
$|Z-Z_0|<\Delta Z$, where the nominal vertex position $Z_{0}$ and
cut parameter $\Delta Z$ values are: $-579.5$ and 5.5 cm, $-579.5$
and 1.5 cm, $-579.5$ and 0.8 cm, $-581.2$ and 0.6 cm for p+p, C+C,
Si+Si and Pb+Pb collisions, respectively. The vertex position in
the transverse $X$, $Y$ coordinates had to agree with the incoming
beam position as measured by the BPD detectors.

In order to reduce the contamination of particles from secondary
interactions, weak decays and other sources of non-vertex tracks,
several track cuts were applied. The accepted particles were
required to have measured points in at least one of the Vertex
TPCs. A cut on the extrapolated distance of closest approach to
the fitted vertex of the particle at the vertex plane was applied
($|d_{X}|<4$~cm and $|d_{Y}|<2$~cm). Moreover the particle was
accepted only when the potential number of points (calculated on
the basis of the geometry of the track) in the detector exceeded
30. The ratio of the number of points on a track to the potential
number of points had to be higher than 0.5 to avoid split tracks
(double counting).


\subsection{Centrality selection}


In order to reduce the effect of fluctuations in the number of
participants (and thus of particle sources $N_{S}$), the
multiplicity fluctuations were analyzed in narrow centrality bins
defined by the energy measured in the VCAL. This procedure
minimizes the variation of the number of projectile participants
$N_{P}^{PROJ}$. Although the number of projectile and target
participants are closely correlated, the number of target
participants cannot be constrained with the NA49 detector. This
leads to a remaining fluctuation in the total number of
participants, i.e. particle sources $N_S$.

For C+C and Si+Si interactions, three and five narrow centrality
bins were selected respectively. In the case of Pb+Pb collisions,
58 narrow centrality bins were chosen. For each centrality bin the
number of projectile participants $N_{P}^{PROJ}$ was estimated by
\begin{equation}
N_{P}^{PROJ}=A-\frac{E_{Veto}}{E_{LAB}}
\label{eqnpproj}
\end{equation}
where $E_{Veto}$ is the energy deposited in the Veto Calorimeter;
$E_{LAB}$ is the energy carried by single projectile nucleons. The
resolution of $E_{Veto}$, which is recalculated into the
resolution of $N_{P}^{PROJ}$, is discussed in Sec.~\ref{s:scaled}.

The effect of smearing of energy carried by spectator nucleons due
to their Fermi momentum was estimated. The effect was found to be
significantly smaller than the resolution of the VCAL and for this
reason can be neglected.

We also studied the contribution of non-spectator particles to the
energy measured by the veto calorimeter~\footnote{The acceptance
of the VCAL at 158A GeV is approximately $\Theta < 0.1$~deg for
neutral particles (e.g. neutrons and photons) and due to the
magnetic field $p>140$~GeV/c, $p_T<0.4$~GeV/c for protons.} with
detailed simulations using VENUS and UrQMD events. The net result
was a reduction of the scaled variance of the multiplicity
distribution by about $0.3$ for central Pb+Pb collisions when the
non-spectator contribution to the VCAL energy is taken into
account. With increasing number of spectators the effect becomes
much smaller due to this effect. No corrections were applied to
the experimental results, nor were the non-spectator contributions
taken into account in the calculations of model predictions.


\subsection{Multiplicity distributions}

\begin{figure}
\begin{center}
\centerline{\includegraphics*[width=8.5cm]{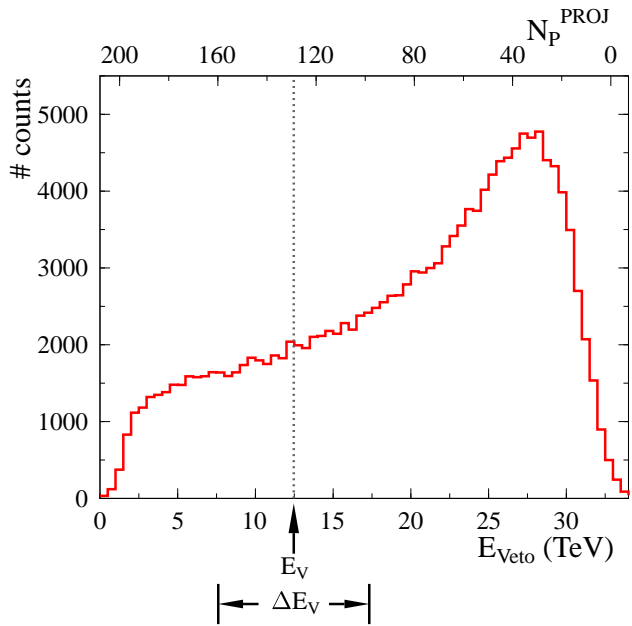}}
\caption{\label{veto}(color online) Distribution of energy
deposited in the Veto Calorimeter for minimum bias Pb+Pb
collisions at $158\,A$~GeV. An example of an $E_{Veto}$ interval
is shown; the interval is determined by its central value $E_{V}$
and the width $\Delta E_{V}$. The width of the histogram bins
corresponds to the width of the $\Delta E_{V}$ intervals used for
the fluctuation analysis. The upper horizontal scale shows the
corresponding values of the number of projectile participants
$N_{P}^{PROJ}$.}
\end{center}
\end{figure}

The multiplicity distribution depends on the selected $E_{Veto}$
interval (its position $E_{V}$ and width $\Delta E_{V}$; see
Fig.~\ref{veto} for definitions) and the kinematic acceptance
selected for the analysis.

\begin{figure*}
\begin{center}
\centerline{\includegraphics*[width=18cm]{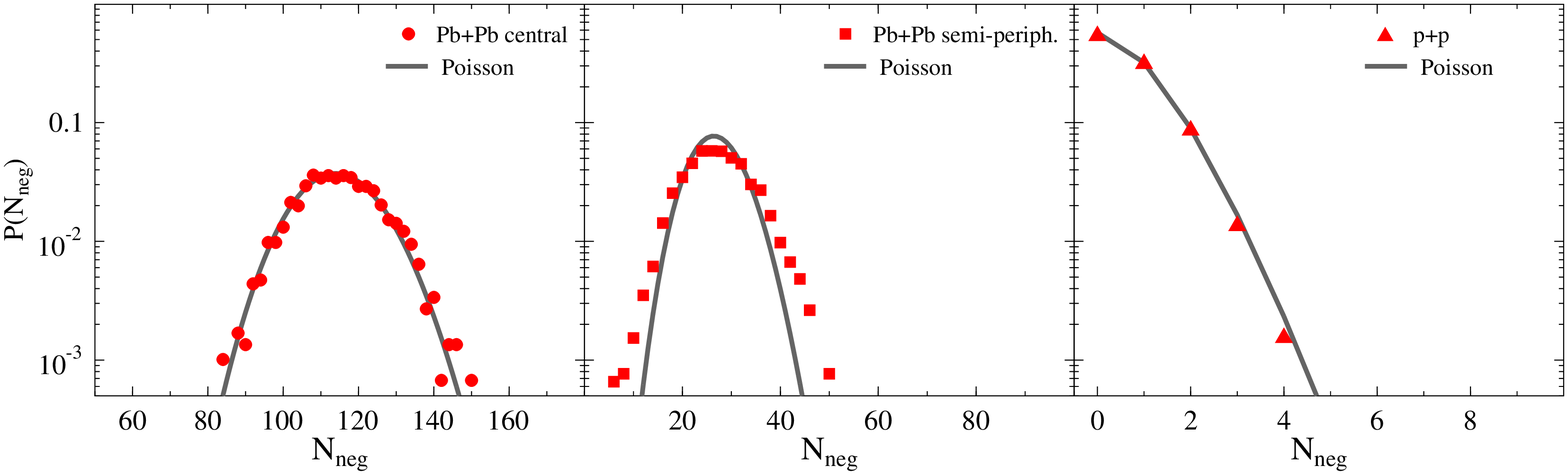}}
\caption{\label{mult3}(color online) Multiplicity distributions of
negatively charged particles obtained for $\Delta E_{V}=500$~GeV
for central Pb+Pb  collisions with number of projectile
participants $N_{P}^{PROJ}=178$ (left panel); semi-peripheral
Pb+Pb collisions with $N_{P}^{PROJ}=39$ (mid-panel) and p+p
interactions with $N_{P}^{PROJ}=1$ (right panel).}
\end{center}
\end{figure*}

In the centrality intervals and acceptance selected for this
analysis, multiplicity distributions show approximately
Poissonian behavior for p+p and central Pb+Pb collisions (see
Fig.~\ref{mult3}). For semi-peripheral collisions the multiplicity
distribution is significantly broader than a Poissonian
distribution.

\begin{figure}
\begin{center}
\centerline{\includegraphics*[width=8.5cm]{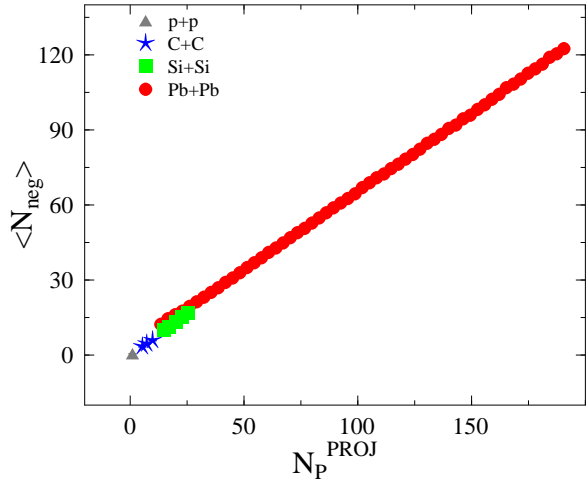}}
\caption{\label{mean_mult}(color online) The uncorrected mean
value of the multiplicity distribution for negatively charged
particles as a function of the number of projectile participants.
(rapidity interval $1.1<y_{c.m.}<2.6$)}
\end{center}
\end{figure}

\begin{figure}
\begin{center}
\centerline{\includegraphics*[width=8.5cm]{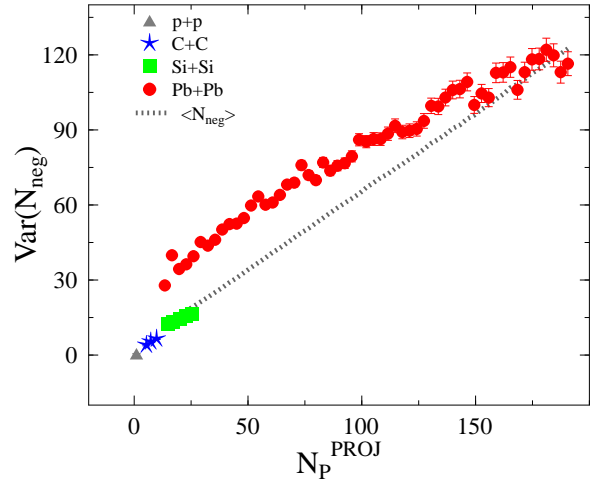}}
\caption{\label{var_mult}(color online) The uncorrected variance
of the multiplicity distribution for negatively charged particles
as a function of the number of projectile participants.}
\end{center}
\end{figure}

In Figs.~\ref{mean_mult} and \ref{var_mult} the measured mean
value and the variance of the multiplicity distributions as a
function of the number of projectile participants $N_{P}^{PROJ}$
are presented. The quantities are {\em not} corrected for the
fluctuations of the number of projectile participants, as
discussed in the next subsection. One can see from these plots
that the mean multiplicity shows an approximately linear
dependence on the number of projectile participants, whereas the
variance of the multiplicity distributions calculated from the
data exceeds the variance of the Poisson distributions.



\subsection{Scaled variance of the multiplicity distribution}
\label{s:scaled}

\begin{figure}
\begin{center}
\centerline{\includegraphics*[width=8.5cm]{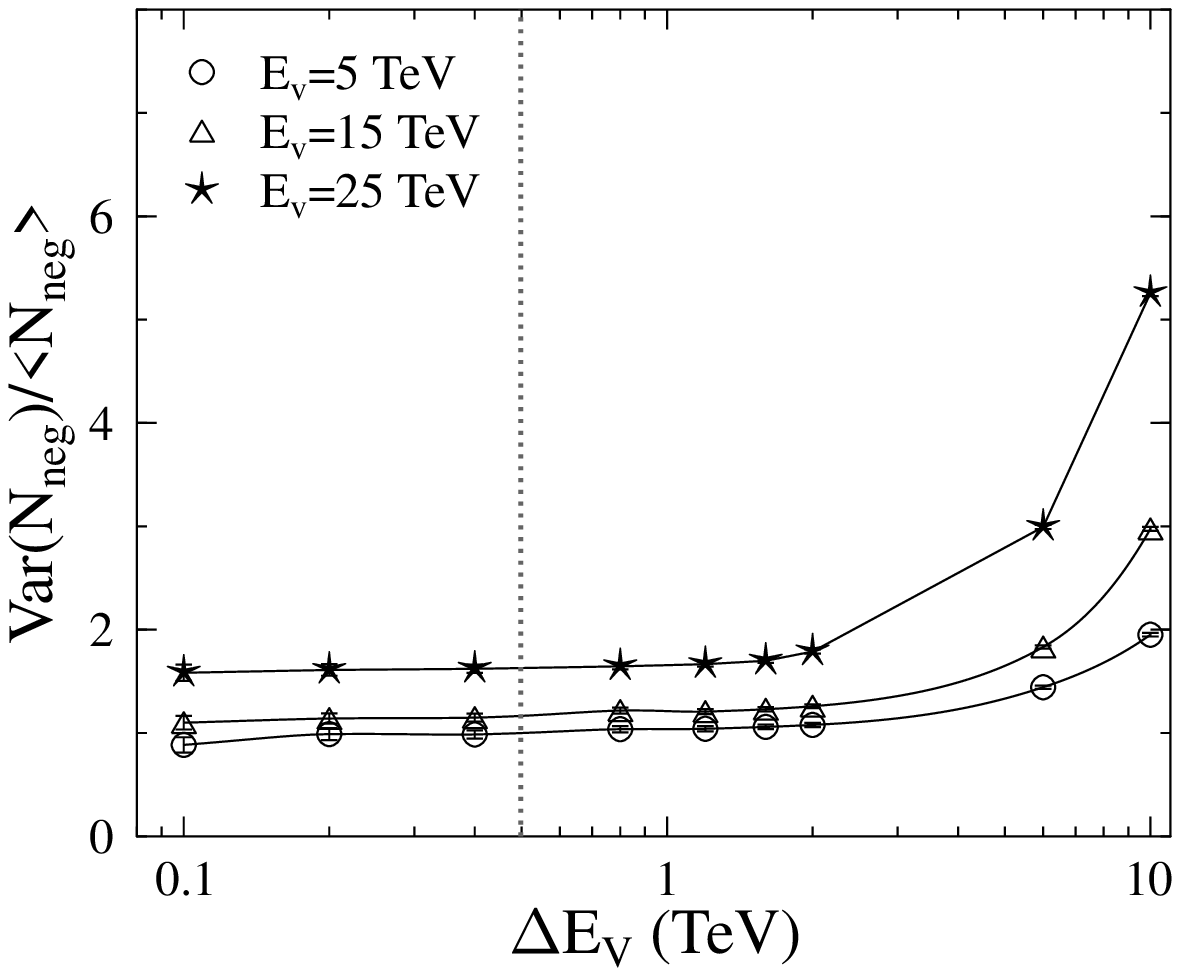}}
\caption{\label{mult_delta}Measured scaled variance of
multiplicity distributions of negatively charged particles for
minimum bias Pb+Pb collisions as a function of the interval width
$\Delta E_{V}$ of the selected VCAL energy for various positions
$E_{V}$ of the interval. The vertical line shows the width of the
interval, which was used for further analysis. See text for
details.}
\end{center}
\end{figure}

The scaled variance of the multiplicity distribution depends,
among other variables, on the width of the energy interval $\Delta
E_{V}$ selected in the VCAL. For very broad $\Delta E_{V}$
intervals the measured scaled variance of the multiplicity
distribution has a large value because of significant fluctuations
in the number of projectile participants $N_{P}^{PROJ}$. Narrowing
the $\Delta E_{V}$ interval results in decreasing fluctuations in
the number of projectile participants and consequently in a
reduction of the scaled variance as shown in
Fig.~\ref{mult_delta}. For $\Delta E_{V}$ smaller than about
$1$~TeV the measured scaled variance of the multiplicity
distribution is almost independent of $\Delta E_{V}$. Note that
even for very small values of $\Delta E_{V}$ the number of
projectile spectators fluctuates due to the finite resolution of
the VCAL. In the following the scaled variance will be calculated
in the $E_{Veto}$ intervals $\Delta E_{V}=500$~GeV (the choice of
the width of the interval is a compromise between minimizing
correction for the finite interval width and sufficient
statistics) and then corrected for fluctuations in the number of
projectile participants due to the finite width of the $E_{Veto}$
interval and the finite resolution of the Veto Calorimeter.

Within the superposition model leading to Eq.~(\ref{sc-var}), the
correction $\delta$ to the scaled variance, which takes into
account the finite value of $\Delta E_{V}$ and the calorimeter
resolution, is calculated as
\begin{equation}
\delta=\frac{\langle N\rangle
\bigl({\rm Var}_{\Delta}\left(E_{Veto}\right)+{\rm Var}_{R}\left(E_{Veto}\right)\bigr)}{\bigl(E_{BEAM}-\langle
E_{Veto}\rangle \bigr)^{2}}, \label{poprawka}
\end{equation}
where ${\rm Var}_{\Delta}\left(E_{Veto}\right)$ is the variance of
$E_{Veto}$ due to the finite width of the $E_{Veto}$ bin, ${\rm
Var}_{R}\left(E_{Veto}\right)$ is the variance of $E_{Veto}$ due
to the VCAL resolution, $\langle E_{Veto}\rangle$ is the mean
value of $E_{Veto}$ in the bin and $E_{BEAM}=158\,A$~GeV is the
total beam energy. ${\rm Var}_{\Delta}\left(E_{Veto}\right)$ was
calculated from the distribution of $E_{Veto}$ energy in a given
interval. ${\rm Var}_R(E_{Veto})=\sigma^2(E_{Veto})$ with
$E_{Veto}$ expressed in GeV was parameterized using calibration
measurements~\cite{MR_PHD} as:
\begin{equation}
\frac{\sigma\left(E_{Veto}\right)}{E_{Veto}}=
\frac{2.85}{\sqrt{E_{Veto}}}+\frac{16}{E_{Veto}} \;.
\label{resolution}
\end{equation}
The resolution of $E_{Veto}$ is easily propagated to the
resolution of $N_{P}^{PROJ}$ using Eq.~(\ref{eqnpproj}).

\begin{figure}
\begin{center}
\centerline{\includegraphics*[width=8.5cm]{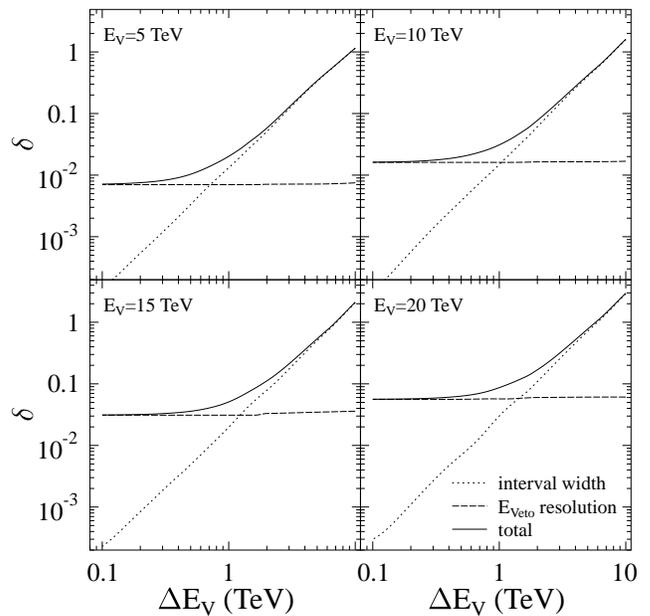}}
\caption{\label{corr}The corrections for Veto Calorimeter
resolution and finite interval width as a function of interval
width $\Delta E_{V}$ for various positions $E_{V}$ of the
interval.}
\end{center}
\end{figure}

\begin{figure}
\begin{center}
\centerline{\includegraphics*[width=8.5cm]{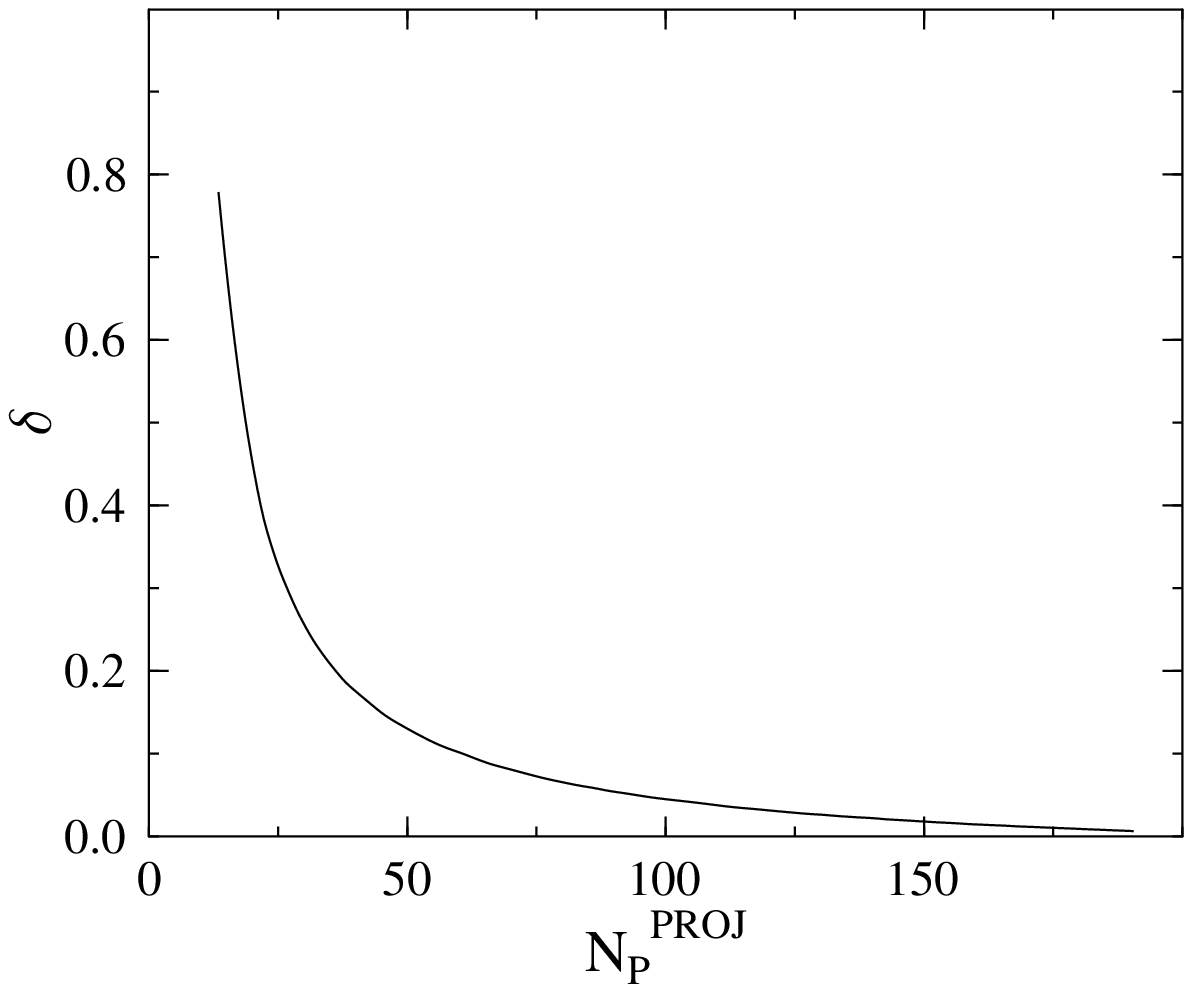}}
\caption{\label{eveto_resolution}The total correction applied to
the measured scaled variance of negatively charged particles as a
function of the number of projectile participants $N_{P}^{PROJ}$.}
\end{center}
\end{figure}

Finally, the corrected scaled variance is calculated as
\begin{equation}
\frac{{\rm Var}\left(n\right)}{\langle
n\rangle}=\frac{{\rm Var}\left(N\right)}{\langle N\rangle}-\delta,
\label{eq_main}
\end{equation}
where ${\rm Var}\left(N\right)$ is the measured variance, $\langle
N\rangle$ the measured mean value of the multiplicity distribution
in a given $E_{Veto}$ bin, and $\delta$ represents the corrections
for fluctuations in the number of projectile participants. The
dependence of $\delta$ on $\Delta E_{V}$ for some values of
$E_{V}$ are presented in Fig.~\ref{corr}, while the total
correction $\delta$ as a function of number of projectile
participants $N_{P}^{PROJ}$ is presented in
Fig.~\ref{eveto_resolution}. Using the same procedure,
corresponding corrections were determined for C+C and Si+Si
collisions. The corrected values of the scaled variance from
Eq.~(\ref{eq_main}) are plotted in Fig.~\ref{varn_all}.

Let us discuss in more detail the physical meaning of the scaled
variance displayed in Fig.~\ref{varn_all} which is the main result
of our study. As already stressed, fluctuations of the number of
projectile participants contribute to the multiplicity
fluctuations of interest. However, the scaled variance of the
multiplicity distribution can be split within the superposition
model according to formula~(\ref{sc-var}), that is into the
contributions coming from a single participant (or equivalently
from a fixed number of participants) and from the varying number
of participants - the first and the second term in
Eq.~(\ref{sc-var}), respectively. Since the correction $\delta$
represents the contribution from the fluctuating number of
projectile participants, the corrected scaled variance shown in
Fig.~\ref{varn_all} describes fluctuations at fixed number of
projectile participants.

If the number of projectile participants uniquely determines the
number of particle sources, which contribute to the rapidity
interval under study, the scaled variance from Fig.~\ref{varn_all}
represents the multiplicity fluctuations of particles coming from
a single source. However, the number of particle sources can
fluctuate even at fixed number of projectile participants. For
example, the number of strings, which decay into particles, is a
random variable in some models. The number of particle sources can
also fluctuate due to the fluctuating number of target
participants which are not observed in the NA49 detector. Thus, we
stress that the scaled variance shown in Fig.~\ref{varn_all}
represents the multiplicity fluctuations at fixed number of
projectile participants and may still contain some contribution
from fluctuations in the number of particle sources.


\subsection{Statistical errors}


In general, the variance of a quantity expressed by the fraction
$\omega=x/y$ is given as
\begin{equation}
{\rm Var}(\omega) \approx
\bigg(\frac{\langle x\rangle}{\langle y \rangle}\bigg)^2
\bigg[\frac{\sigma_{x}^{2}}{\langle x \rangle^2}
+\frac{\sigma_y^2}{\langle y \rangle^{2}}
-\frac{2 {\rm Cov}(x,y)}{\langle x\rangle \langle y \rangle} \Bigg],
\end{equation}
where the symbol $\langle\ldots\rangle$ means an average over
events, $\sigma$ denotes the statistical error of $x$ or $y$, and
${\rm Cov}(x,y)$ is the covariance between $x$ and $y$. The
statistical error of the quantity $\omega$ is given by
$\sigma\left(\omega\right)=\sqrt{{\rm Var}\left(x/y\right)}$.

For the calculation of the statistical errors of the scaled
variance, we have $x={\rm Var}\left(N\right)$, $y=\langle
N\rangle$ and
\begin{gather}
\sigma_{x}=\frac{{\rm Var}\left(N\right)}{\sqrt{L}}\sqrt{2+\gamma_{2}},\;\;
\sigma_{y}=\frac{\sqrt{{\rm Var}\left(N\right)}}{\sqrt{L}},\nonumber\\
{\rm Cov}\left(x,y\right)=\frac{\mu_{3}\left(N\right)}{L},
\end{gather}
where
$\gamma_{2}=\mu_{4}\left(N\right)/\mu_{2}^{2}\left(N\right)-3$ is
the kurtosis, $\mu_{k}\left(N\right)$ is the $k-$th central moment
of the multiplicity distribution, and $L$ is the number of events.

The statistical errors of our results are usually much smaller
than the systematic uncertainties discussed in Sec.~\ref{s:sys}.


\subsection{Stability for event cuts and track selection}


In this section the stability of our results with respect to
variations of the event and track selection criteria is described.
All tests were performed for negatively charged particles produced
in Pb+Pb minimum bias collisions for the centrality bin
corresponding to the number of projectile participants
$N_{P}^{PROJ}=39$.

The event and track selection criteria are designed to reduce the
contamination from the background. To check the stability of the
results obtained for ${\rm Var}\left(n_{neg}\right)/\langle
n_{neg}\rangle$ the cut parameters were varied within reasonable
ranges. In addition, results obtained from the analysis of data
taken at two different magnetic field polarities as well as during
different running periods have been compared.

\begin{figure}
\begin{center}
\centerline{\includegraphics*[width=8.5cm]{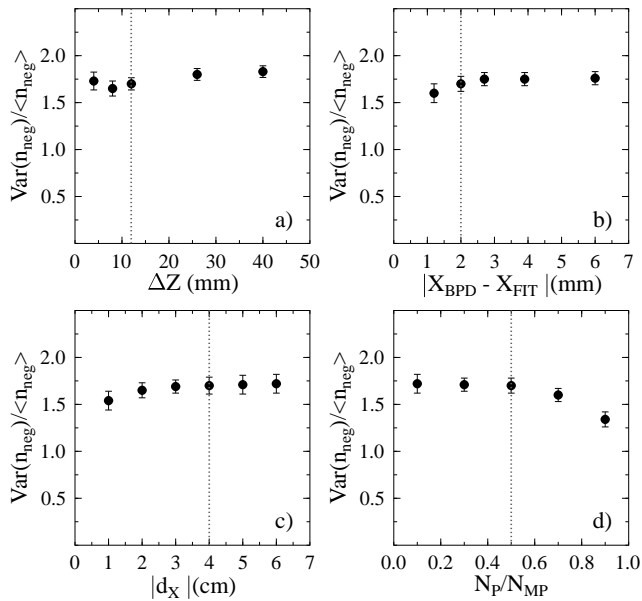}}
\caption{\label{stab}Scaled variance of the multiplicity
distribution of negatively charged particles for the centrality
bin corresponding to the number of projectile participants
$N_{P}^{PROJ}=39$ as a function of: (a) maximally allowed
difference $\Delta Z$ between fitted main vertex and target
position; (b) maximally allowed difference between the transverse
beam position measured by the BPD and the fitted main vertex
position ($|X_{BPD}-X_{FIT}|$) (simultaneously the difference in
$|Y_{BPD}-Y_{FIT}|$ was required to be below $2/3$ of the upper
limit in $|X_{BPD}-X_{FIT}|$); (c) maximally allowed distance
between the reconstructed primary vertex and the track in the
target plane $|d_X|$ (simultaneously the deviation in $|d_Y|$ was
required to be below $0.5\cdot |d_X|$); (d) minimum value of the
ratio of the number of points on the track $N_{P}$ to the
potential number of points $N_{MP}$. The vertical lines indicate
the cuts used to obtain the results.}
\end{center}
\end{figure}

Event cuts are used to reject contamination by non-target
interactions. Fig.~\ref{stab}a shows the scaled variance as a
function of the maximally accepted distance $\Delta Z$ between
fitted and nominal $Z$ position of the primary vertex. As one can
see, the scaled variance of the multiplicity distribution is
stable with respect to a small contamination by non-vertex tracks.

Fig.~\ref{stab}b shows the scaled variance as a function of the
maximally allowed difference of the $X$ and $Y$ position of the
fitted main vertex from the transverse beam position given by the
beam position detectors (cf.~Fig.~\ref{setup}). Also in this case
the scaled variance is stable.

The majority of tracks selected by the track selection criteria
are main vertex tracks and the remaining fraction $(\approx 10\%)$
originates predominantly from weak decays and secondary
interactions with the material of the detector. To estimate the
influence of this contamination on the multiplicity fluctuations,
the maximally accepted distance between the extrapolated track and
the reconstructed primary vertex in the target plane was varied
(see Fig.~\ref{stab}c).

Losses of tracks due to reconstruction inefficiency and track
selection cuts influence the measured multiplicity fluctuations.
In order to estimate this effect, the dependence of ${\rm
Var}\left(n_{neg}\right)/\langle n_{neg}\rangle$ on the ratio of
the number of points on a track to the potential number of points
was determined (see Fig.~\ref{stab}d).

In summary, the values of the scaled variance appear stable with
respect to reasonable variations in the event and track selection
cuts.


\subsection{Systematic uncertainties}
\label{s:sys}

There are several sources of systematic uncertainties of our
results. The systematic error due to the contamination of
non-vertex interactions, tracks from weak decays and secondary
interactions as well as reconstruction inefficiencies and biases
were estimated by varying event and track selection cuts and
simulations as discussed in the previous section. The systematic
error due to the above mentioned effects was estimated to be
$10\%$.

The main source of a possible systematic bias, however, is the
uncertainty in the measurement of the energy deposited in VCAL. To
estimate the influence of this uncertainty on the scaled variance
of the multiplicity distribution the analysis of so-called beam
only events was performed. Here, the Pb beam hits into VCAL
without interactions with target nuclei. The measured resolution
was $\sigma\left(E_{Veto}\right)/E_{Veto}=2.5/\sqrt{E_{Veto}}$
($E_{Veto}$ in GeV)~\cite{MR_PHD}. However, the beam impinges on a
small spot in the VCAL, and the resolution estimated by using beam
only events does not take into account the VCAL non-uniformity
\footnote{The response of the VCAL is different for different
positions of the spectator's hits. It is caused by the
construction of the calorimeter. The four photomultipliers, which
gain the signal, are situated outside of the active area of the
calorimeter sectors. Thus, the amplitude of the signal depends on
the horizontal position, where the signal was generated, because
of the absorption of the ultraviolet light in the scintillator.}.
To estimate it, the correlation between signals deposited in
different sectors of the VCAL were studied. From the result the
resolution (see Eq.~(\ref{resolution})) of the Veto calorimeter
was derived~\cite{MR_PHD}. The systematic error due to the finite
resolution and non-uniformity of the VCAL was estimated to be
smaller than $10\%$ for $N_{P}^{PROJ}>20$ and $25\%$ for
$N_{P}^{PROJ}<20$.

Finally, the systematic error of the results in Pb+Pb collisions
was set to be equal to $10\%$ for $N_{P}^{PROJ}>20$ and $25\%$ for
$N_{P}^{PROJ}<20$. The resulting estimates are shown by the outer
error bars in Fig.~\ref{varn_all}. For semi-central C+C and Si+Si
collisions we assume $10\%$ systematic error (cf.
Figs.~\ref{varn_all_CC}-\ref{varn_all_SiSi}). In p+p interactions
the centrality selection does not apply and the systematic error
was estimated by varying event and track selection cuts yielding
$10\%$ error.


\section{Results}
\label{s:results}


The results discussed in this section refer to the accepted
particles, i.e. particles that are registered by the detector and
pass all kinematic cuts and track selection criteria. The data
cover a broad range in $p_{T}$ ($0.005<p_{T}<1.5$~GeV/c). The
center-of-mass rapidity of accepted particles is restricted to the
interval $1.1 <y_{c.m.} <2.6$ where the azimuthal acceptance given
by Eq.~(\ref{eq_acc}) is large.


\subsection{Multiplicity fluctuations in Pb+Pb collisions}
\label{s:PbPb}

\begin{figure}
\begin{center}
\centerline{\includegraphics*[width=8.5cm]{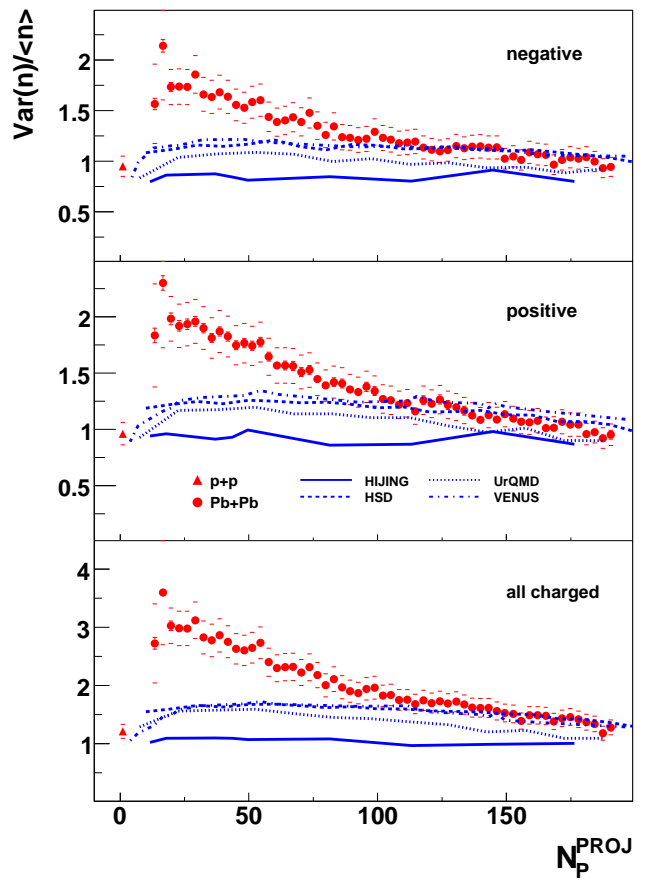}}
\caption{\label{varn_all}(color online) The scaled variance of the
multiplicity distribution for negatively (upper panel), positively
(middle panel) and all (bottom panel) charged particles as a
function of the number of projectile participants $N_{P}^{PROJ}$
compared with model simulations in the NA49 acceptance (HSD and
UrQMD predictions were taken from~\cite{Konchakovski:2005hq}). The
statistical errors are smaller than the symbols (except for the
most peripheral points). The horizontal bars indicate the
systematic uncertainties (the upper one for the second point is
off scale).}
\end{center}
\end{figure}

The corrected scaled variance of the multiplicity distribution for
negatively, positively and all charged accepted particles produced
in minimum bias Pb+Pb collisions as a function of centrality is
shown in Fig.~\ref{varn_all}
(cf.~\cite{Gazdzicki:2004ef,Rybczynski:2004yw}) and compared to
HIJING~\cite{Gyulassy:1994ew}, HSD~\cite{Cassing:1999es},
UrQMD~\cite{Bleicher:1999xi} and VENUS~\cite{Werner:1993uh}
simulations. In the model calculations a realistic simulation of
the determination of the number of projectile participants was
included. The models produce approximately Poissonian multiplicity
distributions independent of centrality. The data points, in
contrast, indicate a strong increase towards peripheral collisions
possibly with a maximum at about $N_{P}^{PROJ}\simeq 30$
projectile participants. We note that by chance the scaled
variance is approximately $1$ for p+p collisions at $158\,A$~GeV,
although the multiplicity distribution in p+p collisions is not
Poissonian both at lower and higher energies (cf.
Sec.~\ref{s:extrapolation} and \cite{Gazdzicki:1990bp}).

The scaled variances for positively and negatively charged
particles are similar. The corresponding values for all charged
particles are larger. Assuming that negatively and positively
charged particles in the experimental acceptance are correlated
with a correlation factor $\rho$, one gets
\begin{equation}
\frac{{\rm Var}\left(n_{ch}\right)}{\langle n_{ch}\rangle}=
\frac{{\rm Var}\left(n_{neg}\right)}{\langle n_{neg} \rangle}
\left(1 + \rho\right)=\frac{{\rm Var}\left(n_{pos}\right)}
{\langle n_{pos}\rangle}\left(1+\rho\right)
\end{equation}
where $n_{neg}$, $n_{pos}$ and $n_{ch}$ are multiplicities of
negatively, positively and all charged particles, respectively,
and, for simplicity, we assumed $\langle n_{neg}\rangle=\langle
n_{pos}\rangle$. As seen, the scaled variance for charged
particles is $1+\rho$ times as large as that for positives or
negatives (see also~\cite{Begun:2004gs}). The positive value of
the correlation factor required by the data ($\rho \approx 0.5$)
may reflect the effect of electric charge conservation.

\subsection{Multiplicity fluctuations in C+C and Si+Si collisions}
\label{s:CC_SiSi}


\begin{figure}
\begin{center}
\centerline{\includegraphics*[width=8.5cm]{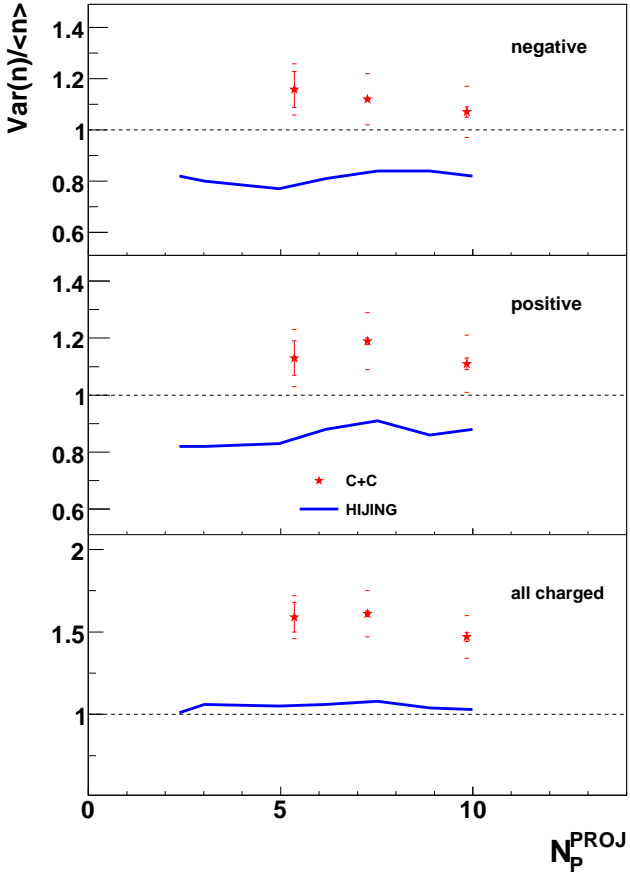}}
\caption{\label{varn_all_CC}(color online) The scaled variance of
the multiplicity distribution of negatively (upper panel),
positively (middle panel) and all (bottom panel) charged particles
produced in semi-central C+C collisions as a function of the
number of projectile participants $N_{P}^{PROJ}$ compared with
HIJING simulations in the NA49 acceptance. The horizontal bars
show the systematic uncertainties while the statistical errors are
represented by the vertical bars. Except for the most peripheral
points the statistical errors are smaller than the symbols.}
\end{center}
\end{figure}

\begin{figure}
\begin{center}
\centerline{\includegraphics*[width=8.5cm]{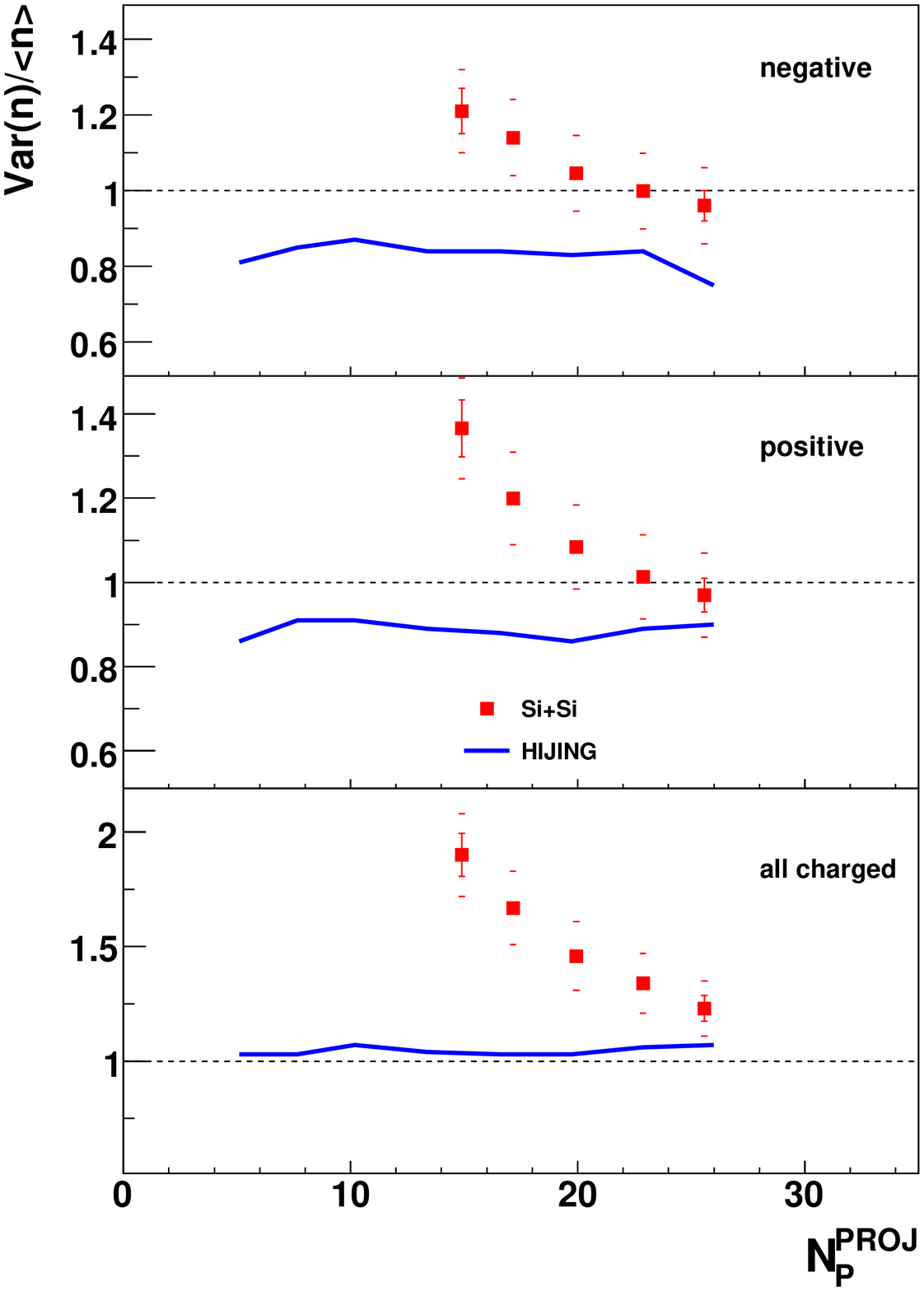}}
\caption{\label{varn_all_SiSi}(color online) The scaled variance
of the multiplicity distribution of negatively (upper panel),
positively (middle panel) and all (bottom panel) charged particles
produced in semi-central Si+Si collisions as a function of the
number of projectile participants $N_{P}^{PROJ}$ compared with
HIJING simulation in the NA49 acceptance. The horizontal bars show
the systematic uncertainties while the statistical errors are
represented by the vertical bars. Except for the most peripheral
points the statistical errors are smaller than the symbols.}
\end{center}
\end{figure}

\begin{figure}
\begin{center}
\centerline{\includegraphics*[width=8.5cm]{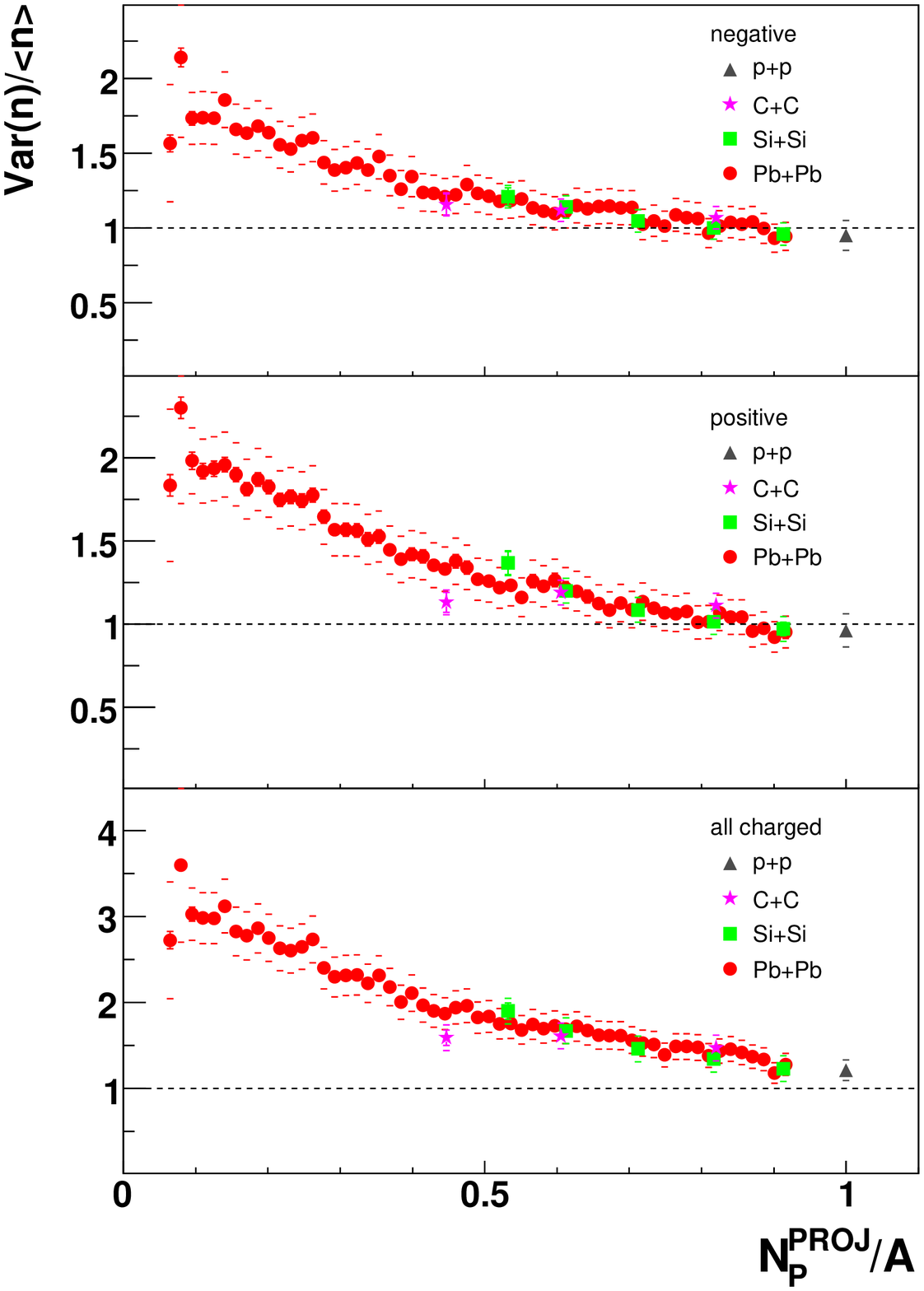}}
\caption{\label{varn_all_NPA}(color online) The scaled variance of
the multiplicity distribution of negatively (upper panel),
positively (middle panel) and all (bottom panel) charged particles
produced in p+p, semi-central C+C, semi-central Si+Si and Pb+Pb
collisions as a function of the fraction of nucleons which
participate in the collision. The statistical errors are smaller
than the symbols (except for the most peripheral points). The
horizontal bars indicate the systematic uncertainties (the upper
one for the second point is off scale).}
\end{center}
\end{figure}

The corrected scaled variances of multiplicity distributions for
negatively, positively and all charged accepted particles produced
in semi-central C+C and Si+Si collisions as a function of
centrality are presented in Figs.~\ref{varn_all_CC} and
\ref{varn_all_SiSi}. The data are compared with results from a
HIJING simulation. The scaled variance for negative and positive
hadrons shows almost Poissonian fluctuations in C+C as well as in
Si+Si collisions, similarly to the observed multiplicity
fluctuations in central Pb+Pb interactions. The multiplicity
fluctuations for all charged accepted particles are much larger
than for the like sign, as also observed in Pb+Pb collisions.
Since data on peripheral C+C and Si+Si collisions are not
available, it is difficult to speculate about the centrality
dependence of the scaled variance. However, it seems that there is
a scaling seen in the variable $f=N_{P}^{PROJ}/A$, the fraction of
nucleons participating in the collision (Fig.~\ref{varn_all_NPA}).
In order to validate this very suggestive scaling data on minimum
bias collisions of C+C and Si+Si would be necessary.


\section{Comparison with results from other experiments}
\label{s:extrapolation}


Assuming that the produced particles are emitted independently in
momentum space, the mean multiplicity in a limited acceptance
$\langle N \rangle$ can be expressed through the mean multiplicity
in the full acceptance $\langle N_{p=1}\rangle$ as $\langle
N\rangle = p \langle N_{p=1}\rangle$, where $p$ is the fraction of
registered particles. The variance of the multiplicity
distribution in a limited acceptance can be written as ${\rm
Var}\left(N\right)=p^{2}{\rm Var}\left(N_{p=1}\right)+\langle
N_{p=1}\rangle p\left(1-p\right)$. The probability that a given
fraction of produced particles is registered is given by the
binomial distribution. Using the above formulas, one easily finds
the scaled variance in a limited acceptance
\begin{equation}
\label{acceptance}
\frac{{\rm Var}\left(N\right)}{\langle N\rangle}=1+p
\bigg(\frac{{\rm Var}\left(N_{p=1}\right)}{\langle N_{p=1}\rangle}
-1 \bigg) \;,
\end{equation}
which depends linearly on the fraction $p$ of registered
particles.

\begin{figure}
\begin{center}
\centerline{\includegraphics*[width=8.5cm]{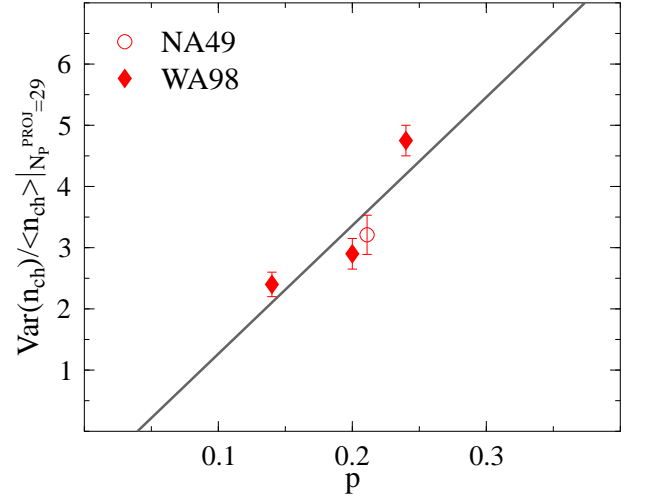}}
\caption{\label{acc_max_rap_pt}(color online) The scaled variance
of the multiplicity distribution of all charged particles produced
in semi-peripheral Pb+Pb collisions as a function of the fraction
of registered particles. The red diamonds correspond to the WA98
data \cite{Aggarwal:2001aa}. The error bars represent the
systematic uncertainties, the statistical errors are smaller than
the symbols. The straight line shows a linear fit to the points.
See text for the details.}
\end{center}
\end{figure}

In order to compare to results obtained by other experiments, the
NA49 results were extrapolated to the full azimuthal acceptance in
the rapidity interval $1.1 <y_{c.m.} <2.6$. The extrapolation was
done by means of Eq.~(\ref{acceptance}) and values of $p$ obtained
from a HIJING simulation which included the NA49 acceptance
filter. Fig.~\ref{acc_max_rap_pt} shows the scaled variance of the
multiplicity distribution of all charged particles for
$N_P^{PROJ}=29$ from the NA49 and WA98~\cite{Aggarwal:2001aa}
experiments plotted versus the fraction $p$ of registered
particles in the NA49 experiment and in the WA98
experiment~\cite{Aggarwal:2001aa}. The latter measured the
multiplicity fluctuations of charged particles produced in Pb+Pb
collisions at $158\,A$~GeV in the central rapidity region. The
predicted linear dependence of Eq.~(\ref{acceptance}) is seen to
agree with WA98 results~\cite{Aggarwal:2001aa}. The WA98
experiment used a centrality determination method which is
different from NA49, i.e. the total transverse energy $E_{T}$ in
the forward rapidity region was used to fix the centrality in the
WA98 experiment. Nevertheless the results of NA49 and WA98 at the
same value of $p$ agree (see Fig.~\ref{acc_max_rap_pt}).

The NA49 detector registers a fraction $p \simeq 0.55$ of all
produced particles. The fraction $p$ of registered particles can
be decreased by cuts in the rapidity distribution. In the case of
p+p interactions, the scaled variance of the multiplicity
distribution was calculated in the following rapidity intervals
$\Delta y$: (1.1,~2.6), (0.85,~3.8), (0.0,~3.8), (-0.4,~3.8) and
(-2.4,~3.8) which correspond to fractions $p$ of registered
particles equal to 0.18, 0.25, 0.41, 0.48 and 0.55, respectively.
Instead of cutting in rapidity one can change the fraction of
registered particles by selecting a range in transverse momentum
$p_T$ of the registered particles. The scaled variance of the
multiplicity distribution was then calculated in the following
$\Delta p_{T}$~(GeV) intervals: (0.0,~0.2), (0.0,~0.4), (0.0,~0.6)
and in the full range corresponding to $p$ equal 0.2, 0.37, 0.41
and 0.55, respectively.

\begin{figure}
\begin{center}
\centerline{\includegraphics*[width=8.5cm]{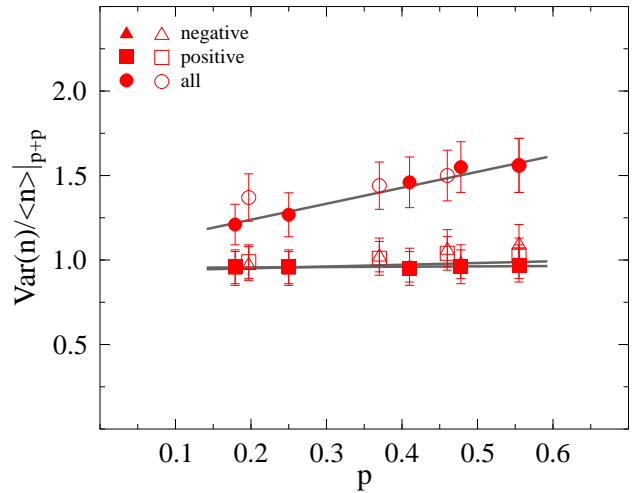}}
\caption{\label{acc_pp}(color online) The scaled variance of the
multiplicity distribution of negatively, positively and all
charged particles produced in p+p collisions as a function of the
fraction of registered particles chosen by change of rapidity
(full symbols) or transverse momentum (open symbols) interval. The
error bars represent the systematic uncertainties, the statistical
errors are smaller than the symbols. The straight lines show the
linear fits to the points.}
\end{center}
\end{figure}

Fig.~\ref{acc_pp} shows the scaled variance of the multiplicity
distribution for negatively, positively and all charged particles
produced in p+p collisions at $158\,A$~GeV as a function of the
fraction $p$ of registered particles chosen by changing the
rapidity and transverse momentum intervals. One can clearly see
the linear dependence described by Eq.~(\ref{acceptance}). The
dependence of the scaled variance on the fraction of registered
particles is independent of the choice of rapidity or transverse
momentum as cut variable.

The rise of the scaled variance of all charged particles with
increasing acceptance can be presumably attributed to the effect
of resonances which predominantly decay into oppositely charged
particles.

\begin{figure}
\begin{center}
\centerline{\includegraphics*[width=8.5cm]{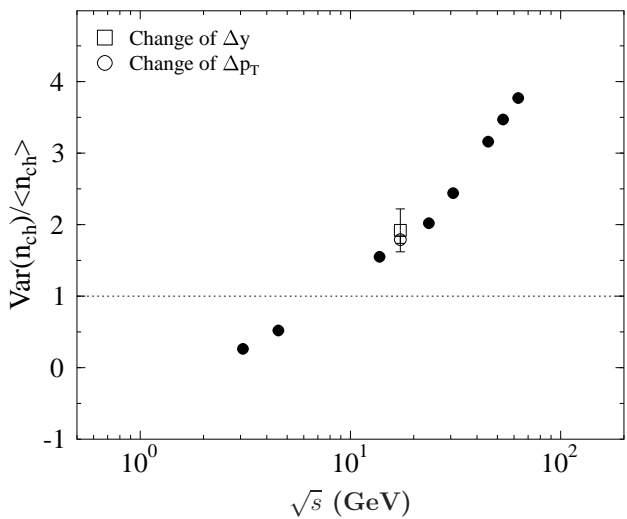}}
\caption{\label{varn_en}The scaled variance of the multiplicity
distribution of charged particles produced in p+p interactions as
a function of the center of mass energy. The full circles show a
compilation of p+p results
(cf.~\cite{Thome:1977ky,Whitmore:1973ri,Whitmore:1976ip}). The
open symbols show our results obtained by the extrapolation by
changing the rapidity $\Delta y$ or transverse momentum $\Delta
p_{T}$ intervals.}
\end{center}
\end{figure}

Multiplicity distributions were measured in p+p interactions in
numerous experiments in a wide range of collision energies. An
excellent compilation of p+p data can be found in
\cite{Thome:1977ky,Whitmore:1973ri,Whitmore:1976ip}.
Fig.~\ref{varn_en} shows the scaled variance of multiplicity
distributions of all charged particles in full phase-space as a
function of the collision energy. The extrapolated result from
NA49 (Eq.~(\ref{acceptance})) is also shown in Fig.~\ref{varn_en}.
The corrected scaled variance of all particles reaches a value of
about 2 and fits well to the other data shown in this plot.


\section{Comparison with models and discussion}
\label{s:models}


Before closing our considerations we discuss possible origins of
the strong dependence of ${\rm Var}(n)/\langle n \rangle$ on the
collision centrality. In this context, it should be stated that
the strong centrality dependence of the scaled variance found in
our experimental study was completely unexpected. We do not know
of any theoretical considerations that would have predicted the
effect. However, after release of our preliminary data
\cite{Gazdzicki:2004ef,Rybczynski:2004yw} several attempts to
explain them have been formulated
\cite{Rybczynski:2004zi,Cunqueiro:2005hx,Brogueira:2005cn,Gazdzicki:2005rr,Konchakovski:2005hq}.
Below the attempts are briefly presented.

We start the discussion with a reminder that the multiplicity
fluctuations presented in this study were observed for a fixed
number of the projectile participants. Although the average
numbers of participants from the projectile and from the target
are approximately equal in the collisions of identical nuclei, the
number of target participants fluctuates even though the number of
projectile participants is kept fixed. We also note that the
multiplicity fluctuations were measured in the forward hemisphere
($1.1 <y_{c.m.} <2.6$). Keeping these remarks in mind, we note
that the large multiplicity fluctuations seen in the forward
rapidity window are caused either by dynamical fluctuations of the
particle production process, which are present even at fixed
number of projectile participants, or the multiplicity
fluctuations in the forward hemisphere result from the
fluctuations of target participant number (the number is not
experimentally fixed) which are transferred to the forward
hemisphere due to an unknown collision mechanism. The mechanism
has to strongly couple the forward and backward hemispheres.

For the purpose of the further discussion, it is useful to divide
the models of nucleus-nucleus collisions into three categories as
proposed in \cite{Gazdzicki:2005rr}. There are {\it transparency
models} where the projectile (target) participants, which are
excited in the course of interaction, mostly contribute to the
particle production in the forward (backward) hemisphere.
Furthermore, there are (rather unrealistic) {\it reflection
models} when the projectile (target) participants mostly
contribute to the backward (forward) hemisphere. Finally, there
are {\it mixing models} in which the projectile and target
participants contribute significantly to both the backward and
forward hemispheres.

In the transparency models, the observed multiplicity fluctuations
can be caused only by the already mentioned dynamical fluctuations
of the production process. In the reflection and mixing models,
the fluctuations of the target participant number contribute to
the multiplicity fluctuations in the forward hemisphere in
addition to possible dynamical fluctuation. The models, which are
represented in Figs.~\ref{varn_all}, \ref{varn_all_CC} and
\ref{varn_all_SiSi}, that is HIJING~\cite{Gyulassy:1994ew},
HSD~\cite{Cassing:1999es}, and UrQMD~\cite{Bleicher:1999xi} all
belong to the transparency class. The results of
VENUS~\cite{Werner:1993uh} are very similar to those of HIJING,
HSD and UrQMD, even though the correlation between the forward and
backward hemispheres is somewhat stronger in VENUS. The model
calculations were performed with proper simulation of the VCAL
response and the NA49 acceptance. The $E_{Veto}$ energy of a model
event was calculated as the energy of the projectile spectators
smeared by a Gaussian distribution with the width given by
Eq.~(\ref{resolution}). The models, which produce approximately
Poissonian multiplicity distributions independently of centrality,
highly underestimate the observed multiplicity fluctuations in
non-central collisions. The models are unable to reproduce even
qualitatively the centrality dependence of the scaled variance.
This remark applies not only to the Monte Carlo models as
HIJING~\cite{Gyulassy:1994ew}, HSD~\cite{Cassing:1999es}, and
UrQMD~\cite{Bleicher:1999xi}, which are based on string excitation
and decay, but to any transparency model which does not assume
correlations among secondary particles. For example, within a
statistical model with fixed volume, where the electric charge is
strictly conserved, the scaled variance of positive or negative
particles varies in the range $0.5-1.0$~\cite{Begun:2004gs}. The
actual value depends on the thermal system's volume and the
acceptance in which the multiplicity fluctuations are observed.

\begin{figure}
\begin{center}
\centerline{\includegraphics*[width=8.5cm]{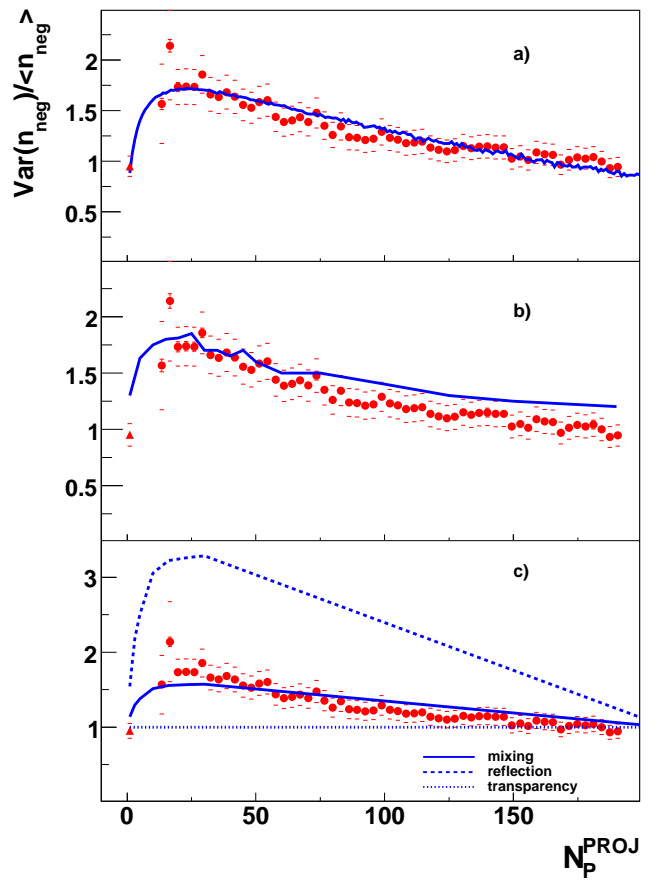}}
\caption{\label{models_1}(color online) The scaled variance of the
multiplicity distribution of negatively charged particles produced
in Pb+Pb minimum bias collisions as a function of the number of
projectile participants. The data are compared to the predictions
of the following models: (a) the model which assumes the
inter-particle correlations caused by the combination of strong
and electromagnetic interactions \cite{Rybczynski:2004zi}, (b) the
percolation model \cite{Cunqueiro:2005hx}, (c) the transparency,
mixing and reflection models \cite{Gazdzicki:2005rr}, see text for
the details.}
\end{center}
\end{figure}

As already noted, the transparency models can produce large
multiplicity fluctuations, if there are built in correlations
among produced particles. The problem can be formulated in a quite
general way.

The average multiplicity of secondaries is roughly proportional to
the number of nucleons participating in the collision $N_{P}$ (cf.
Fig.~\ref{mean_mult}). Assuming that $N_{P}$ is proportional to
the system's volume $V$, we have $\langle N \rangle = \bar\rho \:
V$ with $\bar\rho$ being the constant density of produced
particles. Defining the correlation function $\nu ({\bf r}_1 -
{\bf r}_2) = \nu ({\bf r})$ through the equation $\rho_2 ({\bf
r}_1, {\bf r}_2) = \rho ({\bf r}_1) \: \rho ({\bf r}_2) \big( 1 +
\nu ({\bf r}_1 - {\bf r}_2)\big)$, where $\rho_2 ({\bf r}_1, {\bf
r}_2)$ is the two-particle density, we get the desired formula
\begin{equation}
\label{var-corr1} \frac{{\rm Var}(N)}{\langle N \rangle} =
1 + \bar\rho \int_V d^3r \:\nu ({\bf r}) \;,
\end{equation}
which tells us that the multiplicity fluctuations are related to
the inter-particle correlations. The multiplicity distribution is
Poissonian, if the particles are emitted independently from each
other ($\nu ({\bf r})$ = 0).

It is not difficult to invent a correlation function $\nu ({\bf
r})$ which substituted in Eq.~(\ref{var-corr1}) reproduces the
data shown in Fig.~\ref{varn_all}. Some functions are discussed in
\cite{Rybczynski:2004zi}. The correlation function has to be
positive at small distances (attractive interaction) and negative
at larger distances (repulsive interaction). For Pb+Pb collisions
the sign of the correlation changes at $r \approx 4 \; {\rm fm}$
which corresponds to $N_P \approx 70$ when ${\rm Var}(N)/\langle N
\rangle$ reaches its maximum. For $r \gtrsim (300)^{1/3}\; {\rm
fm} \approx 7 \; {\rm fm}$ the correlation function vanishes.
Possible mechanisms responsible for such a correlation function
were discussed in \cite{Rybczynski:2004zi}. They include: a
combination of attractive and repulsive interaction (the
prediction of this model is shown in Fig.~\ref{models_1}a),
percolation, dipole-dipole interaction, and non-extensive
thermodynamics.

Specific percolation models, which produce strong correlations
among secondaries and thus provide large multiplicity
fluctuations, were discussed in
\cite{Cunqueiro:2005hx,Brogueira:2005cn}. The color strings, which
are stretched in the course of the interaction, form multi-string
clusters. In p+p collisions there are single-string clusters but
in central Pb+Pb collisions the density of strings is so high that
the strings overlap and form one big cluster. The decrease of the
scaled variance as a function of centrality is in this picture
associated with the percolation phase transition, i.e. the
appearance of the large cluster. The prediction of the
model~\cite{Cunqueiro:2005hx} is shown in Fig.~\ref{models_1}b and
reproduces the qualitative behavior of our experimental results.

While the transparency models require strong inter-particle
correlations to comply with our data, the mixing and reflection
models can produce large multiplicity fluctuations in the forward
hemisphere due to the fluctuations of target participant number
which is not fixed in our measurement. In Fig.~\ref{models_1}c we
show the prediction of the reflection and mixing models
\cite{Konchakovski:2005hq}. In the reflection model the target
participants contribute to the forward hemisphere, and thus the
fluctuations of the target participant number produce the
multiplicity fluctuations of secondaries. As seen in
Fig.~\ref{models_1}c, the reflection model strongly overestimates
the observed scaled variance but the mixing models, where the
target as well as projectile participants equally contribute to
the forward and backward hemispheres, approximately agree with the
data. The question arises whether the mixing models properly
represent the dynamics of nucleus-nucleus collisions at
high-energies. A recent theoretical analysis of d+Au collisions at
RHIC \cite{Bialas:2004su} suggests that the contribution of
projectile participants indeed extends to the backward hemisphere
while that of target participants to the forward hemisphere.

In principle, the discussed models can predict multiplicity
fluctuations not only in Pb+Pb collisions but also in p+p, C+C and
Si+Si reactions. It remains to be seen whether they can provide a
consistent description of the fluctuations in all reactions.

Finally we note that the transverse momentum fluctuations measured
in nuclear collisions at $158\,A$~GeV~\cite{Anticic:2003fd} and
quantified by the $\Phi(p_T)$ show a similar centrality dependence
with a maximum located close to that of the scaled variance of the
multiplicity distribution. This behavior of the transverse
momentum fluctuations as a function of collision centrality was
related in a superposition model~\cite{Mrowczynski:2004cg} to the
centrality dependence of the multiplicity fluctuations using the
correlation of average $p_T$ and multiplicity observed in p+p
collisions \cite{Anticic:2003fd}. Moreover, we note that the
percolation string model also describes qualitatively the
centrality dependence of $\Phi(p_T)$~\cite{Ferreiro:2003dw}.


\section{Summary and conclusions}
\label{s:summary}


Multiplicity fluctuations in the forward hemisphere were analyzed
by studying the scaled variance of the multiplicity distributions
in p+p, C+C, Si+Si, and Pb+Pb collisions. A strong centrality
dependence is seen for negatively, positively and all charged
particles. The number of projectile participants was used to
determine the collision centrality. The scaled variance is close
to unity for p+p, C+C, Si+Si and central Pb+Pb collisions. However
it increases significantly towards peripheral Pb+Pb collisions.
The magnitude of the scaled variance is similar for positively and
negatively charged particles and is about $1.5$ times larger for
all charged particles.

The string-hadronic models of nuclear collisions (with or without
color exchange and without string fusion) predict no dependence of
the scaled variance on the number of projectile participants, and
thus the models qualitatively disagree with the data. A maximum of
the scaled variance appears for semi-peripheral collisions when
the forward and backward hemispheres are strongly coupled and
fluctuations of the number of target participants contribute to
the forward rapidity window. The observed scaled variance can also
be reproduced within the models which assume strong dynamic
correlations as, for example, in percolation phase transitions.
\newline

\begin{acknowledgements}

This work was supported by the US Department of Energy Grant
DE-FG03-97ER41020/A000, the Bundesministerium fur Bildung und
Forschung, Germany (06F140), the Virtual Institute VI-146 of
Helmholtz Gemeinschaft, Germany, the Polish State Committee for
Scientific Research (1 P03B 097 29, 1 P03B 121 29,  1 P03B 127
30), the Hungarian Scientific Research Foundation (T032648,
T032293, T043514), the Hungarian National Science Foundation,
OTKA, (F034707), the Polish-German Foundation, the Korea Science
\& Engineering Foundation (R01-2005-000-10334-0) and the Bulgarian
National Science Fund (Ph-09/05).

\end{acknowledgements}

\end{document}